\documentclass[11pt,a4paper]{article}

\usepackage{jheppub}
\usepackage[usenames,dvipsnames]{xcolor}
 
\usepackage{amssymb} 
\usepackage{amsmath}
\usepackage{mathtools}
\usepackage{amsfonts}    
\usepackage{dsfont}
\usepackage{pdfpages}
\usepackage{verbatim}
\usepackage{stmaryrd}
\usepackage{graphicx}
\usepackage{tikz}
\usepackage{pgfplots}
\usetikzlibrary{intersections, pgfplots.fillbetween}
\usetikzlibrary{arrows}
\usetikzlibrary{decorations.pathmorphing}
\usetikzlibrary{ decorations.markings}
\usetikzlibrary{shapes.misc}
\tikzset{cross/.style={cross out, draw=black, minimum size=2*(#1-\pgflinewidth), inner sep=0pt, outer sep=0pt},
cross/.default={1pt}}
\usetikzlibrary{shapes.geometric}
\tikzset{
    partial ellipse/.style args={#1:#2:#3}{
        insert path={+ (#1:#3) arc (#1:#2:#3)}
    }
}

\renewcommand{\figurename}{Fig.}

\newcommand{\dd}{\text{d}}

\newcommand{\bi}{\begin{itemize}}
\newcommand{\ei}{\end{itemize}}
\newcommand{\bea}{\begin{eqnarray}}
\newcommand{\eea}{\end{eqnarray}}
\newcommand{\be}{\begin{equation}}
\newcommand{\ee}{\end{equation}}

 \definecolor{pink}{rgb}{1.0, 0.13, 0.32}

\numberwithin{equation}{section}

\begin{document}

\vspace*{2.5cm}
\begin{center}
{ \LARGE \textsc{Remarks On 2D Quantum Cosmology}}
\\
\vspace*{1.7cm}

Dionysios Anninos,$^{1,2}$ Chiara Baracco,$^1$ and Beatrix M\"uhlmann$^3$ \\ 

\vspace*{0.6cm}
\vskip4mm
\begin{center}
{
\footnotesize
{$^1$Department of Mathematics, King's College London, Strand, London WC2R 2LS, UK \newline
$^2$Instituut voor Theoretische Fysica, KU Leuven, Celestijnenlaan 200D, B-3001 Leuven, Belgium \newline
$^3$Department of Physics, McGill University, Montreal, QC H3A 2T8, Canada  \\
}}
\end{center}

\begin{center}
{\textsf{\footnotesize{
dionysios.anninos@kcl.ac.uk, chiara.baracco@kcl.ac.uk}, beatrix.muehlmann@mcgill.ca} } 
\end{center}

\vspace*{0.6cm}



\end{center}
\vspace*{1.5cm}
\begin{abstract}
\noindent
We consider two-dimensional quantum gravity endowed with a positive cosmological constant and coupled to a  conformal field theory of large and positive central charge. We study  cosmological properties at the classical and quantum level. We provide a complete ADM analysis of the classical phase space, revealing a family of either bouncing or big bang/crunch type cosmologies. At the quantum level, we solve the Wheeler-DeWitt equation exactly.  In the semiclassical limit, we link the Wheeler-DeWitt state space to the classical phase space. Wavefunctionals of the Hartle-Hawking and Vilenkin type are identified, and we uncover a quantum version of the bouncing spacetime. We retrieve the Hartle-Hawking wavefunction  from the disk path integral of timelike Liouville theory. To do so, we must select a particular contour in the space of complexified fields.  The quantum information content of the big bang cosmology is discussed, and contrasted with the de Sitter horizon entropy as computed by a gravitational path integral over the two-sphere.
\end{abstract}
$\,$\newline\newline\newline
\begin{center}{\it Dedicated to the memory of Jim Hartle.}\end{center}

\newpage

\tableofcontents

\section{Introduction}

Theoretically tractable models, at the quantum level, of spacetimes evolving with appreciable time dependence are few and far between. This is particularly so for cosmological spacetimes exhibiting periods of  quasi-de Sitter expansion. The situation stands in sharp contrast to counterpart models for quantum black holes (e.g. \cite{Callan:1992rs,Strominger:1996sh,Maldacena:2016hyu}), systems at criticality (e.g. \cite{Wilson:1971dc,Belavin:1984vu}), magnetic phenomena (e.g. \cite{Ising:1925em,Onsager:1943jn}), strongly coupled gauge theories (e.g. \cite{Schwinger:1962tp,tHooft:1974pnl,Seiberg:1994rs}), and more general interacting quantum field theories (e.g. \cite{Coleman:1974bu,Zamolodchikov:1978xm}). The necessity for such models was already emphasised in early treatments on cosmology.\footnote{For instance in Chapter I of Bondi's 1952 monograph ``Cosmology" \cite{alma990001967900206881} one reads ``to construct mentally a number of different models [...]  each of them interesting and remarkable in its own right, and the question of which of them was the `actual' universe became of lesser interest".} There are valuable exceptions to this rule, including recent developments on de Sitter-like rearrangements/deformations of an underlying AdS/CFT Hilbert space \cite{Coleman:2021nor,Anninos:2017hhn,Susskind:2022dfz,Narovlansky:2023lfz}, de Sitter models with higher-spin gauge symmetries \cite{Anninos:2011ui,Anninos:2017eib,Neiman:2022qfq}, and models of two-dimensional quantum gravity with positive cosmological constant $\Lambda>0$ \cite{CarneirodaCunha:2003mxy,Martinec:2003ka,Martinec:2014uva,Bautista:2015wqy,Anninos:2021ene,Muhlmann:2022duj,Maldacena:2019cbz,Cotler:2019nbi}. 
\vspace{0.2cm}\newline
A reason for the aforementioned dearth may be, in part, that the construction of quantum observables in cosmological spacetimes remains a challenging task  as compared to observables in asymptotically anti-de Sitter or Minkowski spacetimes, as reviewed in \cite{Witten:2001kn,Anninos:2012qw}. In cosmological spacetimes, the Cauchy surface is often compact and observers may be surrounded by cosmological horizons shielding away any asymptotic portions of the spacetime. On the other hand, at least through the incorporation of Euclidean methods, gravitational theories with $\Lambda>0$ permit a semiclassically well-defined gauge-invariant path integral that can be used to compute interesting calculables including Euclidean partition functions \cite{Gibbons:1976ue,Anninos:2020hfj} and certain wavefunctionals \cite{Hartle:1983ai,Maldacena:2002vr,Hertog:2011ky,Anninos:2014lwa,Chakraborty:2023yed}. 
\newline\newline
In this paper we explore models of two-dimensional quantum gravity with $\Lambda>0$ coupled to a conformal field theory of large and positive central charge $c$. It has been long appreciated that these models admit a semiclassical limit permitting de Sitter solutions \cite{Polchinski:1989fn}, and have interesting cosmological properties \cite{Martinec:2003ka,CarneirodaCunha:2003mxy,Martinec:2014uva,Bautista:2015wqy}. Here we would like to pursue these models as precision models of quantum de Sitter space, along the line presented in \cite{Anninos:2021ene,Anninos:2023exn}. We consider Lorentzian features of the model, and the structure of both the semiclassical and exact Hilbert space. Our work  naturally builds on previous work \cite{Fukuyama:1988ae,Polchinski:1989fn,Cooper:1991vg,Teitelboim:1983ux}, and draws particular inspiration from
the work of Martinec and collaborators \cite{CarneirodaCunha:2003mxy,Martinec:2014uva,Martinec:2003ka}. We place an emphasis on path integral methods, semiclassical limits, and provide a careful characterisation of the classical phase space. The models are endowed with a large number of locally propagating degrees of freedom, and exhibit some features of general interest. These include big bang/crunch type classical solutions, a large solution space to the Wheeler-DeWitt equations, a non-trivial form of the de Sitter entropy, a tractable Euclidean path integral with an unbounded conformal mode, and quantum fluctuations of the metric itself. 
\newline\newline
We can contrast these models to the pure de Sitter JT gravity model. The latter contains no locally propagating degrees of freedom, exhibits a non-normaliseable Hartle-Hawking type state \cite{Maldacena:2019cbz,Cotler:2019nbi,Iliesiu:2020zld}, and an infinite two-sphere path integral/de Sitter entropy \cite{Nanda:2023wne}. What is less clear in the models we consider, as compared to the pure de Sitter JT models, as well as more general dilaton gravity models, is how to handle non-trivial topology \cite{Maldacena:2019cbz,Cotler:2019nbi,Cotler:2024xzz,Liouvillestring}.

\subsection*{Outline of the paper}

In section \ref{AdMsec}, we consider the classical problem in the ADM formalism. We derive a set of classical phase space equations and constraints, and discuss the general classical solution space. Previous work includes \cite{Teitelboim:1983ux,Martinec:2014uva}. We emphasize the gravitational perspective, and find an interesting discretum of solutions in the classical phase space, in addition to bounce and big bang/crunch geometries satisfying a two-dimensional version of the cosmic no hair conjecture \cite{Hawking:1981fz,Wald:1983ky}. We note that the linearised perturbations around a de Sitter solution take the form of a tachyonic wave equation whose group theoretic origin is related to the discrete series irreducible representation, as analysed recently in \cite{Anninos:2023lin}.
\newline\newline
In section \ref{LiouvilleQM}, we consider the quantisation of the model in the conformal gauge. The methodology arises from early work relating Liouville theory to two-dimensional quantum gravity \cite{Distler:1988jt,David:1988hj} and the corresponding Schr\"odinger picture that stems from its quantisation (as reviewed for instance in \cite{Ginsparg:1993is,Betzios:2020nry}). We implement both Hamiltonian and momentum constraints throughout our analysis. The Liouville theory at hand is a timelike Liouville theory, and we emphasise that the timelike nature of the Liouville theory leads to a `wrong' sign kinetic term in the Schr\"odinger equation that parallels one of the main features of the Wheeler-DeWitt equation \cite{deWitt}.
\newline\newline
In section \ref{cosmoWF}, we study the exact solutions to the wave equation. In the semiclassical limit, the wavefunctions are contrasted with the classical solution space. Wavefunctions of the Hartle-Hawking \cite{Hartle:1983ai} and Vilenkin \cite{Vilenkin:1986cy} type are identified. It is noted that when the matter fields are in a sufficiently excited state, all wavefunctions must have both a contracting and expanding branch in some regime. We discuss various possible candidate inner products for the quantum theory. 
\newline\newline
In section \ref{HHPI}, we consider the ground state wavefunctions from a Euclidean path integral perspective. We implement a saddle point approximation, and identify candidate saddles that match either the Hartle-Hawking or Vilenkin type states from the exact wavefunction analysis. We study the Euclidean path integral both in the absence of any operator insertions, as well as with an operator insertion of vanishing conformal weight. 
\newline\newline
We end with an outlook and future directions in section \ref{outlook}. An emphasis is placed on the quantum information content encoded in a big bang state as compared to the de Sitter entropy carried by a single static patch. We also discuss an alternative perspective of the disk path integral as a thermal partition function of the static patch in the presence of a timelike boundary. Finally, we remark on the issue of non-trivial topology.




\section{Classical Theory}\label{AdMsec}

In this section, we consider the classical limit of two-dimensional quantum gravity coupled to a conformal field theory (CFT) of large and positive central charge $c$ \cite{Teitelboim:1983ux}. The Lorentzian action of our theory will consist of the combination 
\begin{equation}\label{Sgrav}
S_\text{tot} =  \vartheta \int_{\mathcal{M}} \dd^2x \sqrt{-g} R  - 2\vartheta \int_{\partial\mathcal{M}} \dd x \sqrt{h} K -   \Lambda \int_{\mathcal{M}} \dd^2x \sqrt{-g}  + S_{\text{CFT}} ~.
\end{equation}
The combined first and second terms are topological, and we will restrict our analysis to manifolds of fixed topology. Our gravitational theory includes a positive cosmological constant $\Lambda>0$, and the CFT action is denoted by $S_{\text{CFT}}$. Our spatial slice will have either an $\mathbb{R}$ or $S^1$ topology, and we take our two-dimensional pseudo-Riemannian manifold $\mathcal{M}$ to be either $\mathbb{R}\times S^1$ or $\mathbb{R}\times \mathbb{R}$. The boundary term, with $K$ being the trace of the extrinsic curvature and $h$ the induced metric on the boundary $\partial\mathcal{M}$, ensures we have a well-posed variational problem upon imposing Dirichlet boundary conditions at the boundary of the manifold $\partial{\mathcal{M}}$. Non-trivial topologies are suppressed for large values of $\vartheta$, which we take to be the case. 
\newline\newline
In the absence of matter fields, the theory is classically trivial due to the fact that the Einstein tensor vanishes identically. This is no longer the case once we incorporate matter fields since the classical field equations now depend on the stress tensor $T^{\text{m}}_{\mu\nu}$ of the matter fields as well, and read
\begin{equation}
- \Lambda g_{\mu\nu} + T^{\text{m}}_{\mu\nu} = 0~, \quad\quad T^{\text{m}}_{\mu\nu} \equiv \frac{2}{\sqrt{-g}} \frac{\delta \log Z_\text{CFT}}{\delta g^{\mu\nu}}~,
\end{equation}   
where $-\log Z_{\text{CFT}}$ is the contribution to the effective action of the metric stemming from the CFT. The precise form of $T^{\text{m}}_{\mu\nu}$ will depend on the background metric, rendering the above equation non-trivial. We will now discuss a particular choice for the contribution.

\subsection{Matter contribution}

For a large part of our discussion, we will take the matter CFT to reside in its vacuum sector. As such, instead of dealing directly with the CFT fields, we will deal with their contribution to the gravitational action upon being placed in their vacuum state in a given background metric $g_{\mu\nu}$. Defining the vacuum in a general background, void of any symmetries, is subtle. 
We will define the vacuum to be the Hartle-Hawking state \cite{PhysRevD.28.2960}, which is obtained by path integrating the theory on a Riemannian manifold with a single-boundary. Moreover, we will take the configurations to be residing on a geometry that is close to a two-dimensional de Sitter spacetime (dS$_2$). As we shall see, this is justified in the large $c$ limit, whereby one can perform a saddle point approximation around a classical dS$_2$ solution. 
\newline\newline
From a wavefunctional perspective, states in the Hilbert space are given by functionals, $\Psi_{\text{grav}}[h,X_{\text{b}}]$, of the induced metric $h$  and the matter fields $X_{\text{b}}$ along  a spatial slice. If we imagine fixing the background metric,  $\Psi_{\text{grav}}[h,X_{\text{b}}]$ is given by the Hartle-Hawking state of the matter theory in that particular background. Observables are given by Hermitian functional operators that are subject to diffeomorphism constraints (see \cite{vanderDuin:2024pxb} for a recent discussion). We can obtain an effective description of the gravitational sector by integrating out the matter fields in their Hartle-Hawking state. In this case, and assuming trivial topology, the vacuum contribution is given by the Polyakov action \cite{Polyakov:1981rd}. Ordinarily, this action is studied in Euclidean signature where it reads
\begin{equation}
S_{\text{P}}[g_{\mu\nu}] = \frac{c}{96\pi} \int_{\mathcal{M}} \dd^2 x \sqrt{g} R \nabla_g^{-2} R~,
\end{equation}
with $\mathcal{M}$ taken to be an $S^2$ topology. The non-local nature of $S_{\text{P}}$ reflects the fact that we have integrated out fields in a gapless theory. 
\newline\newline
To continue the Polyakov action to Lorentzian signature, it is convenient to first render it local by introducing an auxiliary field $\sigma$, such that 
\begin{equation}
S_{\text{P}}[\sigma,g_{\mu\nu}] = \frac{c}{96\pi}\int_{\mathcal{M}} \dd^2x \sqrt{g}\left(g^{\mu\nu}\partial_\mu\sigma\partial_\nu\sigma + 2R\sigma\right)~.
\end{equation}
In the above, we have dropped a functional determinant pre-factor, as we are working to leading order at large $c$ when performing our classical analysis. One can now continue the above action to Lorentzian signature, and we find
\begin{equation}
S_{\text{Lor}}[\sigma,g_{\mu\nu}] = -\frac{c}{96\pi}\int_{\mathcal{M}} \dd^2x \sqrt{-g}\left(g^{\mu\nu}\partial_\mu\sigma\partial_\nu\sigma + 2R\sigma\right)~.
\end{equation}
By combining the original terms in (\ref{Sgrav}) with $S_{\text{Lor}}$, we arrive at the gravitational theory of interest, whose action is given by
\begin{equation}
S_{\text{grav}}[g_{\mu\nu}] =  \vartheta \int_{\mathcal{M}} \dd^2x \sqrt{-g} R - 2\vartheta \int_{\partial\mathcal{M}} \dd x \sqrt{h} K -   \Lambda \int_{\mathcal{M}} \dd^2x \sqrt{-g}  + S_{\text{Lor}}[g_{\mu\nu}]~,
\end{equation}
where $K$ is the trace of the extrinsic curvature at the boundary $\partial \mathcal{M}$, and $h$ is the corresponding induced metric. We will now analyse this gravitational theory using the ADM formalism.

\subsection{ADM formalism}

A general two-dimensional metric can be parameterised in the standard Arnowitt-Deser-Misner (ADM) form as follows
\begin{equation}
  ds^2 = -N(t,x)^2\dd t^2 + \omega(t,x)^2\left(\dd x + N^x(t,x) \dd t\right)^2 \:,
\end{equation}
where $N(t,x)$ and $N^x(t,x)$ are the lapse and shift fields, whilst the square of $\omega(t,x)$ is the induced metric on a constant $t$ spacelike slice. We will further take $x$ to be a coordinate on $\mathbb{R}$ or $S^1$. 
\newline\newline
In terms of the ADM quantities, our gravitational action reads
\begin{equation} \label{admaction}
\begin{split}
    S_{\text{grav}}[\omega, N, N^x] = & \int_{\mathcal{M}} \dd t \dd x \biggl\{ N\left[-\Lambda\omega - \frac{c}{96\pi}\frac{(\sigma')^2}{\omega} + \frac{c}{24\pi}\left(\frac{\sigma''}{\omega} - \frac{\omega'\sigma'}{\omega^2}\right)\right] + \\ 
    & + \frac{1}{N}\biggl[\frac{c}{96\pi}\omega(\dot{\sigma}-N^x\sigma')^2 + \frac{c}{24\pi}(\dot{\sigma}-N^x\sigma')(\dot{\omega}-(N^x\omega)')\biggr] \biggr\}~,
\end{split}
\end{equation}
where the dot signifies a $t$-derivative and the prime signifies an $x$-derivative. For the time being, we have discarded the topological term, as we will be interested in a fixed topology. We will be mostly interested in the case of a compact spatial surface such that there are no spatial boundary terms. When the spatial slice is non-compact, we will take configurations to fall off sufficiently fast so that there are no boundary contributions to the conserved charges at spatial infinity either.\footnote{One could also consider relaxing the latter condition, as was considered in \cite{Anninos:2010zf,Kelly:2012zc}.}
\newline\newline
The conjugate momenta associated to the various fields are given by
\begin{eqnarray}
    \pi_\sigma &=&\frac{c}{48\pi N}\omega(\dot{\sigma} - N^x\sigma') + \frac{c}{24\pi N}(\dot{\omega} - (N^x\omega)') ~,  \label{momentum2} \\ 
    \pi_\omega &=&  \frac{c}{24\pi N}(\dot{\sigma} - N^x\sigma')~, \label{momentum3} \\
    \pi_{N}  &=& 0~, \quad  \pi_{N^x}  = 0~.  \label{momentum1}
\end{eqnarray}
The above satisfy the Poisson bracket algebra, for instance $\{\sigma(x) , \pi_\sigma(y)\} = \delta(x-y)$ 
and similarly for the other phase space quantities. 
Finally, one has the Hamiltonian and momentum constraints stemming from the equations of motion of $N$ and $N^x$
\begin{eqnarray} 
    \mathcal{H} &=& \frac{24\pi}{c}\pi_\omega \pi_\sigma - \frac{6\pi\omega}{c}\pi_\omega^2 + \Lambda\omega + \frac{c \, \sigma'^2}{96\pi\omega} - \frac{c}{24\pi}\left( \frac{\sigma''}{\omega} - \frac{\omega'}{\omega^2}\sigma' \right) = 0~, \label{hamconstraint} \\
 \label{momconstraint}
  {\Pi} &=&  \sigma' \pi_\sigma -\omega  \pi_\omega'= 0~.
\end{eqnarray}  
We note that $\Pi$ generates tangential diffeomorphisms along the constant $t$ slice.
\newline\newline
In addition to our gauge constraints, $\sigma$ must satisfy its equation of motion
\begin{equation}\label{sigmaeom}
-\nabla^\mu \partial_\mu \sigma + R = 0~.
\end{equation}
Having obtained our constraint equations, we can now analyse them in a judiciously chosen gauge. As our gauge choice, which we refer to as the Weyl gauge, we take the metric to be conformally equivalent to a two-dimensional Minkowski metric. In order to impose the Weyl gauge, we choose $N^x(t,x) = 0$ and $N(t,x) = \omega(t,x)$, leading to the metric
\begin{equation}\label{weylgauge}
ds^2 = \omega^2(t,x) \left( -\dd t^2 + \dd x^2 \right)~.
\end{equation}
The residual gauge symmetries, upon fixing the Weyl gauge, are given by left and right-moving real valued maps
\begin{equation}
x^+ \to f(x^+)~, \quad x^- \to g(x^-)~, \quad\quad\quad x^\pm \equiv t \pm x~,
\end{equation}
which can be reabsorbed in a rescaling
\begin{equation}
\omega^2(x^+,x^-) \to \omega^2\left(f(x^+),g(x^-)\right) f'(x^+)g'(x^-)~.
\end{equation}
Projecting onto a constant $t$ slice, the residual gauge freedom reduces to the set of tangential spatial diffeomorphisms. 
\newline\newline
Upon fixing the Weyl gauge, one can explicitly solve for $\sigma$ in (\ref{sigmaeom}) to find
\begin{equation}
\sigma(t,x) = -2 \log \omega(t,x) + f_\sigma(x^+) + g_\sigma(x^-)~, 
\end{equation}
with $f_\sigma$ and $g_\sigma$ being smooth functions. If we set  both $f_\sigma$ and $g_\sigma$  to zero, we arrive at the classical constraint equations
\begin{eqnarray}
    \mathcal{H} &=& -\frac{c}{24\pi}\frac{3\omega'^2-2\omega\omega''+\dot{\omega}^2}{\omega^3} + \Lambda\omega  = 0 ~,\label{weylhamconst} \\ 
    \Pi &=& \frac{c}{12\pi}\frac{-2\omega'\dot{\omega}+\omega \dot{\omega}'}{\omega^2} = 0~. \label{weylmomconst}
\end{eqnarray}
The above are the classical constraints ensuing from the gauge invariance under the two diffeomorphisms. The constraint equation (\ref{weylhamconst}) ensures independence in the choice of Cauchy slice. The constraint equation (\ref{weylmomconst}) ensures that the configuration space is invariant under those diffeomorphisms tangential to the Cauchy slice. 
\newline\newline
We can also obtain the remaining Hamilton's equation for $\omega$, namely
\begin{equation}
{\delta_\omega H} = - \dot{\pi}_\omega~, \quad\quad\quad\text{with}\quad\quad\quad H \equiv \int \dd x \left( N \mathcal{H} + N^x \Pi \right)~. 
\end{equation}
Fixing the conformal gauge, imposing the on-shell condition on $\sigma$, and implementing the Hamiltonian constraints,  we can arrange the above Hamilton equation   to the form 
\begin{equation}\label{Heom}
-\ddot{\varphi}(t,x) + \varphi''(t,x) = -\frac{24\pi \Lambda}{c} e^{2\varphi(t,x)}~,
\end{equation}
where we have defined $e^{\varphi(t,x)} \equiv \omega(t,x)$. More succinctly,
\begin{equation}\label{Rconst}
R = \frac{48\pi\Lambda}{c}~.
\end{equation}
This is perhaps the closest analogue to an Einstein field equation one can imagine in two spacetime dimensions. It implies that all solutions are diffeomorphic to dS$_2$ in a local neighbourhood of points. Equation (\ref{Heom}) will be identified with the Liouville equation of motion later on.

\subsection{Classical solutions}

Given the phase space equations, we can now discuss their simplest solutions.
\newline\newline
\textbf{Planar dS$_2$.} It is straightforward to see that the constraint equations in the Weyl gauge are solved by the configuration
\begin{equation}\label{planar}
\omega_{\text{p}}(t,x) = -\sqrt{\frac{c}{24\pi\Lambda}}\frac{1}{t}~, \quad\quad \sigma_{\text{p}}(t,x) = -2 \log \omega_{\text{p}}(t,x)~,   \quad\quad t \in (-\infty,0)~,
\end{equation}
where $x\in\mathbb{R}$, and the subscript denotes that this is the planar solution. The minus sign is selected to have a forward pointing time coordinate with positive $\omega_\text{p}(t,x)$. The corresponding phase space momenta read
\begin{equation} \label{planarmomenta}
    \pi_{\sigma_\text{p}} = 0~, \quad\quad \pi_{\omega_\text{p}} = -\sqrt{\frac{c\Lambda}{6\pi}}~.
\end{equation}
Provided $x\in\mathbb{R}$, the solution (\ref{planar}) corresponds to a two-dimensional de Sitter metric in planar coordinates and Ricci scalar $R = \tfrac{48\pi\Lambda}{c}$, namely
\begin{equation}\label{pdS2}
ds^2 = {\frac{c}{24\pi\Lambda}}  \left(\frac{-\dd t^2+\dd x^2}{t^2}\right)~,
\end{equation}
with $t\in (-\infty,0)$ and $x\in\mathbb{R}$. The spacetime curvature becomes parameterically small in the large $c$ limit, which is the dimensionless parameter governing the semiclassical approximation. 
\newline\newline
Taking $\varphi(t,x)  = \log \omega_{\text{p}}(t,x) +  \delta \varphi(t,x)$ and expanding the Hamilton equations of motion (\ref{Heom}) to linear order in the perturbation yields
\begin{equation}
 -\delta \ddot{\varphi}(t,x)  + \delta \varphi''(t,x)   = - \frac{2}{t^2} \delta \varphi(t,x)~.
\end{equation}
The above equation is that of a free tachyonic scalar in a dS$_2$ background of mass squared minus two in units of the de Sitter length. Such tachyonic modes have appeared in the two-dimensional de Sitter literature \cite{Anninos:2023lin,Pethybridge:2024qci,Loparco:2024ibp} and it has been noted that they transform in the discrete series unitary irreducible representation of the $SO(1,2)$ dS$_2$ isometry group. Interestingly, this is also the representation furnished by the linearised metric in four-dimensional de Sitter space (see for instance \cite{Basile:2016aen,Sun:2021thf}). Due to the additional constraints of the gravitational theory, though tachyonic, the perturbation $\delta\varphi$ does not indicate any pathological behaviour. It is not a truly independent degree of freedom. 
\newline\newline
\textbf{Global dS$_2$.} One can also try to find the dS$_2$ solution in global coordinates. Upon inspection, one finds that the global dS$_2$ metric
\begin{equation}\label{globalds2}
\omega_{\text{g}}(t,x) = \sqrt{\frac{c}{24\pi\Lambda}}\frac{1}{\cos t}~, \quad\quad\quad t \in \left(-\frac{\pi}{2},\frac{\pi}{2}\right)~, 
\end{equation}
does not solve the simplified Hamiltonian constraint (\ref{weylhamconst}). At first, this sounds surprising as the global and planar dS$_2$ metrics are diffeomorphic. The reason for the seeming discrepancy is that in order to obtain the global dS$_2$ solution, we can no longer impose the simplifying condition $f_\sigma = g_\sigma = 0$. Instead, we must consider the more general form 
\begin{equation}
\sigma_{\text{g}}(t,x) = -2\log\omega_{\text{g}}(t,x)+f_\sigma(x^+)+g_\sigma(x^-)~.
\end{equation}
Upon further assuming that $\omega(t,x) = \omega(t)$, we are led to a differential equation governing $f_\sigma$ and $g_\sigma$ which is solved by $f_\sigma(x) = g_\sigma(x) =   \log \cos^{-2} \tfrac{x}{2}$. The momentum constraint is satisfied by this ansatz. Some algebra then reveals the following form for the Hamiltonian constraint
\begin{equation}
   -\frac{c}{24\pi}\frac{\dot{\omega}^2}{\omega^2} - \frac{c}{24\pi} + \Lambda\omega^2 = 0 ~\label{weylhamconst2}~.
\end{equation}
\\ Comparing to the previous form (\ref{weylhamconst}), we see that there is a shift between the two. The shift becomes subleading in the regime where $\Lambda\omega^2 \gg \frac{c}{24\pi}$ which we can view as the late time limit of an expanding spacetime. The above equation indeed permits (\ref{globalds2}) as a solution, and the phase space momenta now read
\begin{equation} \label{globalmomenta}
    \pi_{\sigma_\text{g}} = \frac{c}{48\pi}\left(\tan\frac{t-x}{2} + \tan\frac{t+x}{2}\right)~, \quad\quad \pi_{\omega_\text{g}} = -\sqrt{\frac{c\Lambda}{6\pi}}\frac{\cos x \sin t}{\cos t + \cos x}~.
\end{equation}
The physical metric now becomes
\begin{equation}\label{gdS2}
ds^2 = {\frac{c}{24\pi\Lambda}}\left( \frac{-\dd t^2+\dd x^2}{\cos^2{t}}\right)~,
\end{equation}
with $x \sim x + 2\pi$ parameterising an $S^1$ spatial slice. The planar and global dS$_2$ metrics (\ref{pdS2}) and (\ref{gdS2}) are related by a (complexified) Weyl preserving gauge transformation with $f(x^+) = e^{i x^+}$ and $g(x^-) = e^{-i x^-}$. 
\newline\newline
Finally, upon taking $t \to -i\tau$ in (\ref{gdS2}), and extending $\tau \in \mathbb{R}$, one finds the round metric on the two-sphere. The full Euclidean solution is given by
\begin{equation}
\omega_{S^2}(\tau,x) =  \sqrt{\frac{c}{24\pi\Lambda}}\frac{1}{\cosh \tau}~, \quad \tau \in \mathbb{R}~,
\end{equation}
along with 
\begin{equation}
\sigma_{S^2}(\tau,x) = -2 \log \omega_{S^2}(\tau,x) - 2 \log \left( \frac{\cosh\tau+\cos x}{2} \right)~.
\end{equation}
The poles of the two-sphere lie at $\tau = \pm \infty$, and $x \sim x+2\pi$. We note that there is a singular point at $(\tau,x) = (0,\pm \pi)$. In Lorentzian signature the singular point extends along two null directions $t \pm x = \pm \pi$, leading to the singular values of $\sigma_g(t,x)$. Recalling (\ref{sigmaeom}), we note that these are coordinate singularities that don't appear in the physical Ricci scalar. Nonetheless, we would like our functions to be single valued as we now discuss. 
\newline\newline
\textbf{A discretuum of solutions.} It is interesting to note that, in addition to (\ref{gdS2}), it seems that one has an infinite family of smooth solutions to the constraint equations (\ref{weylhamconst2}). In particular, we may consider that the periodicity of $x$ takes an arbitrary real value \cite{Epstein:2018wfh,Anninos:2019oka}.\footnote{The infinite configuration space is also present in four-dimensional gravity with $\Lambda > 0$, upon noting that dS$_2\times S^2$ is a solution to the field equations. Of all these configurations, the one with $x \sim x+2\pi$ is distinguished by permitting a smooth Euclidean configuration.} Requiring, however, that these configurations be globally well-defined, given the periodic structure of $f_\sigma(x) = g_\sigma(x) =   \log \cos^{-2} \tfrac{x}{2}$, enforces that the periodicity be $x\sim x+2\pi n$ with $n\in \mathbb{Z}^+$ --- a discrete subset of the original continuum. 
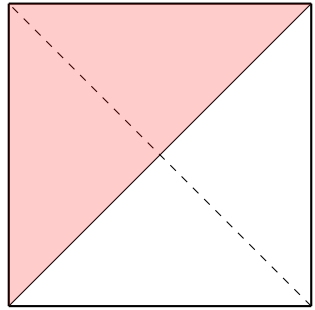
\begin{figure}[ht]
\centering
\begin{tikzpicture}
\draw[thick, name path =A] (0,0) -- (4,0);
\draw[thick, name path =B] (4,0) -- (4,4);
\draw[thick, name path =C] (4,4) -- (0,4);
\draw[thick, name path =D] (0,4) -- (0,0);
\draw[name path =F] (0,0) -- (4,4);
\draw[dashed, name path =G] (4,0) -- (0,4);
\tikzfillbetween[of=F and C]{red, opacity=0.2};
\end{tikzpicture}
\caption{Penrose diagram of global de Sitter space (square) with the shaded region indicating the planar patch (\ref{planar}). For dS$_2$, each point in the square corresponds to an $S^0$, and thus two-points in the physical spacetime. The two asymptotia are given by the horizontal boundaries of the Penrose diagram.}
\end{figure}

\subsection{Adding a matter energy eigenstate} 

As a final remark, we note that adding a matter energy eigenstate of energy $E_{\text{m}}$ modifies the Hamiltonian constraint (\ref{weylhamconst2}) to
\begin{equation}
   -\frac{c}{24\pi}\frac{\dot{\omega}^2}{\omega^2} + \Lambda\omega^2 + E_{\text{m}}  - \frac{c}{24\pi} = 0 ~\label{weylhamconst3}~.
\end{equation}
The energy eigenstate stems from the existing conformal field theory matter fields, which are excited above their ground state (whose Casimir energy on a spatial circle of size $2\pi$ is $-\tfrac{c}{24\pi}$). The variable $E_{\text{m}}$ denotes the energy of the state above the Casimir energy.  In a general curved spacetime, for which there is no timelike Killing vector, the matter Hamiltonian will not admit time independent eigenvalues, so the meaning of $E_{\text{m}}$ might seem unclear. The way we deal with this is by first going to a conformal frame where the conformal fields perceive a flat cylindrical metric, and place them in an energy eigenstate of energy $E_{\text{m}}$ as  measured in this frame. 
\newline\newline
Equation (\ref{weylhamconst3}) admits three distinct types of solution to the equation of motion. For $E_{\text{m}}< \tfrac{c}{24\pi}$ the solutions resemble the global dS$_2$ metric, and have a bounce like structure going from past infinity to future infinity. The energy in this case controls the size of the waist in the bouncing solution 
\begin{equation}\label{bounde}
\omega_{\text{bounce}}(t,x) =   \frac{\sqrt{{ \frac{c}{24\pi\Lambda}-\frac{E_{\text{m}}}{\Lambda}}}}{\cos \left( t \sqrt{1-\frac{24\pi E_{\text{m}}}{c}} \right) }~, \quad\quad t\in  \frac{1}{\sqrt{1-\frac{24\pi E_{\text{m}}}{c}}} \left(-\frac{\pi}{2},\frac{\pi}{2}\right)~,
\end{equation}
with $x\sim x+2\pi$, and
\begin{equation}
\sigma_{\text{bounce}}(t,x) = -2\log\omega_{\text{bounce}}(t,x)+ \log \cos^{-2} \tfrac{t+x}{2}+ \log \cos^{-2} \tfrac{t-x}{2}~.
\end{equation}
At $E_{\text{m}} = \tfrac{c}{24\pi}$ our solution becomes a spatial quotient of the planar slicing of dS$_2$ (\ref{planar}): 
\begin{equation}
    \omega_{\rm quotient}(t,x)= \pm \sqrt{\frac{c}{24\pi\Lambda}}\frac{1}{t}~ ,\quad\quad t\in \mathbb{R}^{\pm}~.
\end{equation}
For $E_{\text{m}} > \tfrac{c}{24\pi}$, we instead have solutions which end in the past or future in a Milne type singularity. The  solution to (\ref{weylhamconst3}) is now given by
\begin{equation}\label{bigbang}
\omega_\pm(t,x) = \pm \frac{\sqrt{{\frac{E_{\text{m}}}{\Lambda} - \frac{c}{24\pi\Lambda}}}}{\sinh \left( t \sqrt{\frac{24\pi E_{\text{m}}}{c}-1} \right) }~, \quad\quad t\in \mathbb{R}^\pm~,
\end{equation}
with $x\sim x+2\pi$, and
\begin{equation}
\sigma_\pm(t,x) = -2\log\omega_\pm(t,x)+ \log \cos^{-2} \tfrac{t+x}{2}+ \log \cos^{-2} \tfrac{t-x}{2}~. 
\end{equation}
The Milne singularity resides at either $t = \pm \infty$, where the proper size of the spatial circle vanishes. In the case of a past singularity, corresponding to $\omega_-(t)$ with $t \in \mathbb{R}^-$, the dynamical evolution is followed by a period of eternal expansion with late time behaviour given by the planar dS$_2$ solution (\ref{planar}) with $x\sim x+2\pi$. This is a version of the cosmic no hair conjecture \cite{Hawking:1981fz,Wald:1983ky} -- the features of the initial condition are washed out at late times. 
\newline\newline
These different solutions are the cosmological analogues \cite{CarneirodaCunha:2003mxy,Martinec:2003ka} of the  hyperbolic, parabolic,  and elliptic  solutions of classical Liouville theory as described, for instance, in \cite{Seiberg:1990eb}. The situation rhymes, imperfectly, with the higher dimensional situation where a sufficiently excited configuration in gravity with $\Lambda>0$ ceases to be asymptotically de Sitter both in the past and future \cite{Friedrich:1986qfi,Andersson:2002nr}. In our case, the cosmological solutions satisfy (\ref{Rconst}), and as such are locally diffeomorphic to dS$_2$. However, they are globally distinct. For instance, if we rescale the bouncing cosmology (\ref{bounde}) to take the form (\ref{gdS2}) of global dS$_2$, we will note that the periodicity of $x$ is no longer $2\pi$. The energy of the matter fields thus has a physical effect on the geometry.

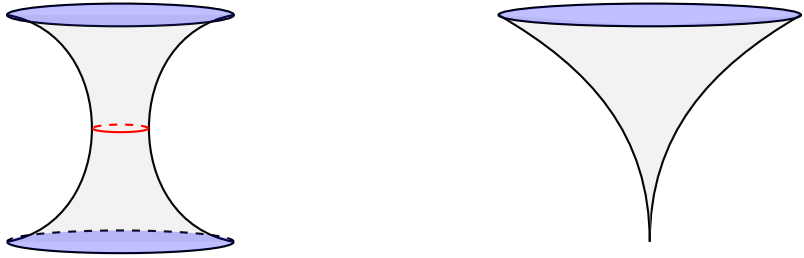
\begin{figure}[ht]
\centering
\begin{tikzpicture}
\draw[thick] (0,3) ellipse (1.5 and 0.15);
\def\xR{1.5}; 
\def\yR{0.15};  
\draw [dashed,domain=0:180,thick] plot ({\xR*cos(\x)}, {\yR*sin(\x)}); 
\draw [domain=-180:0,thick] plot ({\xR*cos(\x)}, {\yR*sin(\x)});
\draw[thick, name path=C] (7,0) to[out=90,in=210] (9,3);
\draw[white, name path=D] (7,0) -- (7,3);
\draw[thick, name path=G] (7,0) to[out=90,in=-30] (5,3) ;
\fill[gray, opacity=0.1] (7,2.85) to (7,0) to[out=90,in=-30] (5,3) ;
\fill[gray, opacity=0.1] (7,2.85) to (7,0) to[out=90,in=210] (9,3);


\draw[thick] (7,3) ellipse (2 and 0.15);
\fill[blue, opacity=0.25] (7,3) ellipse (2 and 0.15);

\draw[thick, name path=A] (1.5,0) to[bend left=80, looseness=1.3] (1.5,3);
\draw[thick, name path=B] (-1.5,0) to[bend right=80, looseness=1.3] (-1.5,3);
\tikzfillbetween[of=A and B]{gray, opacity=0.1};
\fill[blue, opacity=0.25] (0,3) ellipse (1.5 and 0.15);
\fill[blue, opacity=0.25] (0,0) ellipse (1.5 and 0.15);

\def\XR{.37}; 
\def\YR{0.05};  
\draw [dashed,domain=0:180,red,thick] plot ({\XR*cos(\x)}, {1.5+\YR*sin(\x)}); 
\draw [domain=-180:0,red,thick] plot ({\XR*cos(\x)}, {1.5+\YR*sin(\x)});

\end{tikzpicture}
\caption{Left: For $E_{\rm m} < \frac{c}{24\pi}$ we have bouncing cosmologies with two exponentially expanding regions in the far past and far future. The red slice at $t=0$ is the minimal waist of length square $\frac{2\pi}{\Lambda}(\frac{c}{24\pi} - E_{\rm m})$. Right: For $E_{\rm m} > \frac{c}{24\pi}$ we have solutions which end in the past or future in a Milne type singularity. Here we show the case of a past singularity. }
\end{figure}

%

\section{Liouville Quantisation} \label{LiouvilleQM}

In this section, following the treatment of two-dimensional quantum gravity as a Liouville theory \cite{Ginsparg:1993is}, we examine a quantum version of the classical ADM formulae (\ref{weylhamconst}) and (\ref{weylmomconst}). To do so, it is illuminating to perform the field redefinition $\omega(t,x) = e^{\varphi(t,x)}$ such that the classical constraints (\ref{weylhamconst}) and (\ref{weylmomconst}) read
\begin{eqnarray}\label{HamL}
- \frac{c}{24\pi} \left( \dot{\varphi}^2 +  {\varphi'}^2  - 2 \varphi'' \right) + \Lambda  e^{2 \varphi}  &=& 0~, \\ \label{MomL}
\dot{\varphi} \varphi' - \dot{\varphi}'&=& 0~.
\end{eqnarray}
In what follows, the above constraint equations will be understood in the language of Liouville theory and we will then quantise the theory from this perspective \cite{Martinec:2003ka,Martinec:2014uva,Bautista:2015wqy}.

\subsection{Anomaly action \& quantum Liouville}

We begin by identifying (\ref{HamL}) and (\ref{MomL}) in terms of the  stress tensor of the anomaly action that appears upon placing a two-dimensional conformal field theory on a  metric of the form $e^{2\varphi}\tilde{g}_{\mu\nu}$, whose action in Lorentzian signature reads \cite{Polyakov:1981rd}
\begin{equation}\label{SL}
S_{\text{anom}} = \frac{c}{24\pi} \int \dd^2x \sqrt{-\tilde{g}} \left( \tilde{g}^{\mu\nu} \partial_\mu \varphi \partial_\nu\varphi +   \tilde{R} \varphi - \frac{24\pi \Lambda}{c} e^{2\varphi} \right)~,
\end{equation}
where $\tilde{R}$ is the Ricci scalar of $\tilde{g}_{\mu\nu}$. 
\newline\newline
The equations of motion for the above theory on a Minkowski background, where $\tilde{g}_{xx} = -\tilde{g}_{tt}=1$ and $\tilde{g}_{tx}=0$, are given by (\ref{Heom}) which we reproduce below for convenience
\begin{equation}
-\ddot{\varphi}(t,x)  +\varphi''(t,x)=- \frac{24\pi \Lambda}{c} e^{2\varphi(t,x)}~.
\end{equation}
The classical stress tensor of the above action on a Minkowski background is given by
\begin{eqnarray}
T_{tt} &=&  - \frac{c}{24\pi}\left(\dot{\varphi}^2  +\varphi'^2 -2\varphi'' \right)  + \Lambda e^{2\varphi}~, \\
T_{tx} &=&  \frac{c}{12\pi}\left(\dot{\varphi} \varphi' - \dot{\varphi}' \right)~, 
\end{eqnarray}
which we immediately  recognise  as the Hamiltonian (\ref{HamL}) and momentum (\ref{MomL}) densities. 
\newline\newline
Early work \cite{David:1988hj,Distler:1988jt} on two-dimensional quantum gravity indicates that, upon fixing the Weyl gauge, one can describe the exact quantum theory in terms of a timelike Liouville conformal field theory.\footnote{The status of the non-perturbative existence of timelike Liouville theory, which is a non-unitary theory with a `wrong' sign kinetic term, remains somewhat inconclusive. In what follows, as in \cite{Anninos:2021ene}, we consider the theory in semiclassical limit at large $c$, including quantum corrections about the dS$_2$ saddle. Evidence for its existence can be found in \cite{Ribault:2015sxa}.} In Lorentzian signature, the timelike Liouville action reads
\begin{equation}\label{tSL}
S_{\text{tL}} =  \frac{1}{4\pi} \int \dd^2x \sqrt{-\tilde{g}} \left( \tilde{g}^{\mu\nu} \partial_\mu \varphi \partial_\nu\varphi + q \tilde{R} \varphi - 4\pi  \Lambda e^{2 \beta \varphi} \right)~.
\end{equation}
where we have defined $q\equiv 1/\beta-\beta$. The parameter $\beta$ is given by
\begin{equation}
\beta \equiv \frac{\sqrt{c-1}-\sqrt{c-25}}{2 \sqrt{6}}~,
\end{equation}
and is fixed by requiring that the above action is a conformal field theory of central charge
\begin{equation}\label{ctL}
c_{\text{tL}} = 26-c  = 1- 6\left(\frac{1}{\beta} - \beta \right)^2~,
\end{equation}
at least to all orders in the quantum perturbative expansion at small $\beta$ \cite{Anninos:2021ene}. For large $c$, whereby $\beta \approx \sqrt{{6}/{c}}$, we  retrieve (\ref{SL}) upon rescaling $\varphi \to \varphi/\beta$. In our current field variables, the physical metric is given by
\begin{equation}\label{wg2}
g_{\mu\nu}  = e^{2\beta\varphi} \tilde{g}_{\mu\nu}~,
\end{equation}
and the fixed reference metric $\tilde{g}_{\mu\nu}$ is often referred to as the fiducial metric in the literature.  When $\tilde{g}_{\mu\nu}$ is taken to be a two-dimensional Minkowski space, the quantum transformation properties of $\varphi$ under a coordinate transformation $x^+ \to f(x^+)$, and $x^- \to g(x^-)$ are given by \cite{Seiberg:1990eb}
\begin{equation}\label{phitransf}
\varphi(x^+,x^-) \to \varphi(f(x^+),g(x^-)) + \frac{q}{2} \log \partial_+ f(x^+) \partial_- g(x^-)~.
\end{equation}
In the semiclassical limit $\beta\to0^+$, the above reduces to the conformal invariance of the classical timelike Liouville action. The small correction is due to quantum effects.
\newline\newline
In the classical limit, $\beta \to 0^+$, we have that $q=1/\beta$ and the stress tensor stemming from the timelike Liouville action (\ref{tSL})  reads
\begin{equation} \label{stressenergytensor}
 T^{(\text{tL})}_{\mu\nu}  =  \frac{1}{4\pi}\biggl[\tilde{g}_{\mu\nu} \tilde{g}^{\alpha\beta}\partial_\alpha\varphi \partial_\beta\varphi - 2\partial_\mu\varphi \partial_\nu\varphi- 2q\Bigl(-\tilde{\nabla}_\mu\partial_\nu\varphi + \tilde{g}_{\mu\nu}\tilde{\nabla}^\alpha\partial_\alpha\varphi \Bigr)\biggr] - \Lambda \tilde{g}_{\mu\nu}e^{2 \beta \varphi}~.
\end{equation}
The trace of the stress tensor, in the classical limit, is given by
\begin{equation}
\tilde{g}^{\mu\nu} T^{(\text{tL})}_{\mu\nu} =  -\frac{\tilde{R}}{4\pi \beta^2}~,
\end{equation}
where we have used the equations of motion governing $\varphi$. 
We note that the stress tensor, in the classical limit, already captures a significant part of the trace anomaly. The complete trace anomaly follows from a one-loop correction, as shown in \cite{Anninos:2021ene}, and the corrected prefactor of $\tilde{R}$ becomes $\tfrac{c_{\text{tL}}}{24\pi}$ with $c_{\text{tL}}$ as in (\ref{ctL}).  

\subsection{Cosmological Hilbert space}

From the perspective of an underlying gravitational theory, the timelike Liouville action $S_{\text{tL}}$ is not quite complete. In particular, the full gravitational theory must be independent of any Weyl rescalings of the fiducial metric $\tilde{g}_{\mu\nu}$, as these constitute a redundancy of the Weyl gauge (\ref{weylgauge}). This invariance is restored upon incorporating the contributions from the $\mathfrak{b}\mathfrak{c}$-ghost theory of central charge $c_{\text{ghost}}=-26$, and the two-dimensional matter conformal field theory of central charge $c$. The $\mathfrak{b}\mathfrak{c}$-ghost theory is built in a standard fashion from anti-commuting fields $\mathfrak{b}$ and $\mathfrak{c}$ of holomorphic scaling dimensions $\Delta_{\mathfrak{c}} =-1$ and $\Delta_{\mathfrak{b}}=2$, as well as their anti-holomorphic counterpart. The matter theory is taken to be a unitary theory with a discrete spectrum. Altogether, the complete gravitational theory in the Weyl gauge is a conformal field theory with a net vanishing conformal anomaly \cite{Distler:1988jt,David:1988hj} such that 
\begin{equation}\label{ctot}
c_{\text{tL}} + c - 26 = 0~.
\end{equation}
Moreover, physical states in the Hilbert space must lie in the BRST cohomology associated with the BRST charge $\mathcal{Q}_{\text{B}}$.
The physical Hilbert space $\mathcal{H}_{\text{phys}}$ is a subset of the tensor product of the ghost, matter, and Liouville Hilbert spaces $\mathcal{H}_{\text{gh}} \otimes \mathcal{H}_{\text{m}} \otimes \mathcal{H}_{\text{tL}}$. Due to (\ref{ctot}) the physical Hilbert space is a subset of states of a non-anomalous conformal field theory, and both observables and states alike must be Weyl and conformally invariant.
\newline\newline
In what follows, we will consider states in $\mathcal{H}_{\text{phys}}$ whose factor $\mathcal{H}_{\text{gh}}$ is given by the $\mathfrak{b}\mathfrak{c}$-ghost vacuum $|0\rangle_{\text{gh}}$ carrying conformal weights $(-1,-1)$.\footnote{In the case of a minimal model coupled to two-dimensional quantum gravity it was found by Lian-Zuckerman \cite{Lian:1991gk} that the BRST cohomology is significantly richer in the ghost sector, and involves states of arbitrary ghost number. For this to happen, the matter CFT itself must harbor null states. We will assume throughout our analysis that the matter CFT has no such null states --- it is unitary with large, positive central charge, and a discrete spectrum. On the other hand, as for its spacelike counterpart, it is believed that timelike Liouville theory has a collection of degenerate operators (see for instance (2.23) of \cite{Bautista:2020obj}) which may play a role in the solvability of the theory.} The states in the matter Hilbert space $\mathcal{H}_{\text{m}}$ are organised in terms of Virasoro primary states $|\Delta_{\text{m}},\bar{\Delta}_{\text{m}}\rangle$, and labeled by their conformal weights $(\Delta_{\text{m}}, \bar{\Delta}_{\text{m}})$. Given the matter theory is a unitary conformal field theory with $c>1$, there is an infinite number of such Virasoro primaries. At least for the trivial ghost sector $|0\rangle_{\text{gh}}$, the matter Virasoro descendants are gauge equivalent to the Virasoro primaries, via a non-anomalous conformal map, and do not yield physically independent states. The timelike Liouville Hilbert space $\mathcal{H}_{\text{tL}}$  consists of primaries  $|\Delta_P,\bar{\Delta}_P\rangle$ labeled by $\Delta_P = \bar{\Delta}_P$. Normaliseable vertex operators $\mathcal{V}_\alpha = e^{2\alpha \varphi}$ in the timelike Liouville Hilbert space have \cite{Harlow:2011ny,Bautista:2019jau}
\begin{equation}\label{LDelta}
\Delta_P = \bar{\Delta}_P = \alpha(q+\alpha) = -\frac{q^2}{4}+P^2~ ,\quad\quad \text{with}\quad \alpha \equiv - \frac{q}{2}+P~, \quad\quad P \in \mathbb{R}~.
\end{equation}
The conformal dimension is invariant under $\alpha \rightarrow - q-\alpha$ and as such the normaliseable vertex operators $\mathcal{V}_\alpha$ and $\mathcal{V}_{-q-\alpha}$ are identified with each other.
Additionally we have degenerate operators with (see for instance \cite{Ribault:2015sxa})
\begin{equation}\label{eq: degenerate fields}
    \alpha_{m,n} =- \frac{m-1}{2}\beta+\frac{n-1}{2}\beta^{-1}~,
\end{equation}
where $m$ and $n$ are integer valued. Viewed as a theory of quantum gravity these non-normaliseable states  play an important role in the theory. An example of this is the state created by acting with the cosmological operator $ \mathcal{V}_\beta = e^{2\beta\varphi}$ whose conformal weights are $(1,1)$. As we will see below, a priori for timelike Liouville the operators $\mathcal{V}_{\alpha_{m,n}}$ and $\mathcal{V}_{-q-\alpha_{m,n}}$ need not be identified. 
\newline\newline
Given the structure of the pre-Hilbert space, we can now impose the quantum constraints stemming form the underlying gauge-fixed gravitational theory. In the Weyl gauge (\ref{wg2}), the constraints amount to the physical Hilbert space, as well as the space of physical observables, being both conformally and Weyl invariant with respect to transformations of $\tilde{g}_{\mu\nu}$. As such, on a trivial topology, no physical quantity depends on the arbitrary nature of the fiducial metric $\tilde{g}_{\mu\nu}$. At the level of states $|\Psi\rangle$ in the pre-Hilbert space $\mathcal{H}_{\text{gh}} \otimes \mathcal{H}_{\text{m}} \otimes \mathcal{H}_{\text{tL}}$,  the constraints impose that
\begin{equation}\label{qc}
\left( L^{\text{tot}}_0+\bar{L}^{\text{tot}}_0 - 2 \right) |\Psi\rangle = 0 \quad\quad \text{and} \quad\quad \left( L^{\text{tot}}_0 - \bar{L}^{\text{tot}}_0 \right)|\Psi\rangle = 0~,
\end{equation}
where $L^{\text{tot}}_0$ and $\bar{L}^{\text{tot}}_0$ are the zero-modes of the Virasoro generators of the matter and Liouville sector. The above equations are the quantum analogue of the classical Hamiltonian and momentum constraint equations, (\ref{hamconstraint})  and (\ref{momconstraint}), of the previous section. As a consequence of the constraints, taking $|\Psi\rangle  = |\Delta_{\text{m}},\bar{\Delta}_{\text{m}}\rangle \otimes |\Delta_P,\bar{\Delta}_P\rangle \otimes |0\rangle_{\text{gh}}$, we have
\begin{equation}
\Delta_P + \bar{\Delta}_P + \Delta_{\text{m}} +  \bar{\Delta}_{\text{m}} - 1 - 1 = 0 \quad\quad \text{and} \quad\quad \Delta_{\text{m}} = \bar{\Delta}_{\text{m}}~.
\end{equation}
The above constraints express how the timelike Liouville field can gravitationally dress the matter fields to render the state diffeomorphism invariant. The dimension of the physical Hilbert space $\mathcal{H}_{\text{phys}}$ remains infinite dimensional. Nonetheless, it is worth pointing out that there are a finite set of states in a fixed $\Delta_{\text{m}} = \bar{\Delta}_{\text{m}}$ sector.
\newline\newline
The situation we have described is somewhat similar to the treatment of the string worldsheet \cite{Polchinski:1998rq}. However, as we will see in the next section, there are some crucial differences due to the timelike nature of the Liouville field.
\subsection{Wheeler-DeWitt wave equation}

 What we are left to do, to make contact with the ADM analysis of the previous section, is to express the constraint equation (\ref{qc}) in the language of field operators acting on wavefunctionals. As for the classical analysis, we will consider the flat and cylindrical fiducial metrics separately.
\newline\newline
{\textbf{Flat fiducial metric.}}  We will take our quantum Hamiltonian to be the time-time component of the net stress tensor operator, $T_{tt}$, of timelike Liouville theory, coupled to the $\mathfrak{b}\mathfrak{c}$-ghosts and CFT matter, all placed on a metric $\tilde{g}_{\mu\nu}$. We will further represent the Liouville sector of the quantum state $|\Psi\rangle$ in terms of a wavefunctional of the profile, $\varphi(x)$, of the Liouville field on a spatial Cauchy surface. The matter and ghost sectors are taken to be in the state $|\Delta_{\text{m}},\bar{\Delta}_{\text{m}}\rangle \otimes |0\rangle_{\text{gh}}$ with $\Delta_{\text{m}}=\bar{\Delta}_{\text{m}}$.
Taking $\tilde{g}_{\mu\nu}$ to be the Minkowski metric, $\tilde{g}_{xx} = -\tilde{g}_{tt}=1$ and $\tilde{g}_{tx}=0$ with $x\in\mathbb{R}$, the quantum constraints (\ref{qc}) acting on wavefunctionals  $\Psi[\varphi(x)]$  are then given by
\begin{eqnarray}
\biggl[\frac{\delta^2}{\delta \varphi(x)^2} - \frac{1}{4\pi^2}\left(   {\varphi'}^2  - 2 q \varphi'' \right) +  \frac{1}{\pi}{\Lambda}  e^{2 \beta \varphi} + \frac{2{\Delta}_{\text{m}}-2}{2\pi^2}  \biggr]  \Psi[\varphi(x)]  &=& 0~, \label{QH} \\
\biggl[ \varphi'(x) - q\partial_x \biggr] \frac{\delta}{\delta \varphi(x)}\Psi[\varphi(x)] &=& 0~. \label{HamM}
\end{eqnarray}
It is interesting to note that the kinetic term in (\ref{QH}) has the `wrong' sign. This parallels the situation for the ordinary Wheeler-DeWitt equation \cite{deWitt} where certain configurations in the space of all three-metrics also have a wrong sign. The appearance of a wrong sign is intimately tied to the matter CFT having a large positive central charge. Had we considered instead coupling gravity to a matter CFT with $c\le 1$, as reviewed in \cite{Ginsparg:1993is,Betzios:2020nry}, there would be no such wrong sign kinetic term for the gravitational sector and the Weyl mode would be captured by an ordinary spacelike Liouville theory. This again indicates \cite{Polchinski:1989fn,Anninos:2021ene} that a good two-dimensional model for higher-dimensional Euclidean gravity with $\Lambda>0$ requires coupling quantum gravity with $\Lambda>0$ to a two-dimensional CFT with $c\gg 1$. 
\newline\newline
{\textbf{Cylindrical fiducial metric.}} In order to study the wavefunction as a function of the proper length of the boundary, it is convenient to take the fiducial  metric $\tilde{g}_{\mu\nu}$ to be that of a cylinder, with metric 
\begin{equation}
d\tilde{s}^2 = -\dd t^2+\dd x^2~, \quad\quad x\sim x+2\pi~.
\end{equation}
In this case, the Hamiltonian constraint reads
\begin{equation}\label{HamCyl1}
\biggl[\frac{\delta^2}{\delta \varphi(x)^2} - \frac{1}{4\pi^2}\left(   {\varphi'}^2  - 2q\varphi'' \right) + \frac{1}{\pi}{\Lambda}  e^{2 \beta \varphi} - \frac{1}{2\pi^2}\frac{q^2}{2} + \frac{2{\Delta}_{\text{m}}-2}{2\pi^2} \biggr]  \Psi[\varphi(x)] = 0~,
\end{equation}
where we have defined $q\equiv \beta^{-1} - \beta$. One might have expected a constant shift in the Hamiltonian from the anomalous Schwarzian transformation law of the stress-tensor upon mapping the plane to the cylinder. However, the condition (\ref{ctot}) that the total central charge vanishes implies the absence of such a term. Instead, the constant shift $ -\tfrac{q^2}{4}$ in the continuous spectrum of conformal weights (\ref{LDelta}) has been separated from the continuous part of the spectrum.
This is the quantum counterpart of the constant shift between the classical Hamiltonians (\ref{weylhamconst}) and (\ref{weylhamconst2}). At large values of $\varphi$ the two wave-equations (\ref{QH}) and (\ref{HamCyl1}) are approximately equal. Finally, the momentum constraint remains as in (\ref{HamM}). 

\section{Cosmological Wavefunctions}\label{cosmoWF}

Given the quantum constraint equations (\ref{HamCyl1}) and (\ref{HamM}), we can now proceed to analyse their corresponding solutions. The momentum constraint entails that physical states are invariant under diffeomorphisms tangential to the spatial surface. This allows us to further select a gauge condition enforcing the induced metric on the spatial slice to take a constant value $\varphi(x) = \varphi_0$, such that the proper length of the boundary of the physical metric is given by $\ell = 2\pi e^{\beta\varphi_0}$.

\subsection{Wave equation}\label{waveeqn}

Upon going to the gauge $\varphi(x) = \varphi_0$, where
\begin{equation}
\varphi_n \equiv \frac{1}{2\pi}\int_0^{2\pi} \dd x \, e^{-i n x} \, \varphi(x)~, \quad\quad n \in \mathbb{Z}~,
\end{equation}
our wave equation reduces to an ordinary looking Schr\"odinger equation, which can be cast in the form 
\begin{equation}\label{HamCyl}
\biggl(  \frac{\dd^2}{\dd\varphi_0^2} + \lambda^2 e^{2 \beta \varphi_0}  \biggr) \Psi[\varphi_0]=   \varepsilon_{\text{m}}^2 \,  \Psi[\varphi_0]~, 
\end{equation}
where we have defined
\begin{equation}\label{mattercond}
\varepsilon_{\text{m}}^2 \equiv 2\left( \frac{q^2}{2} + {2-2\Delta_{\text{m}}} \right)~ \quad\quad \text{and} \quad\quad \lambda^2 \equiv 4\pi  {\Lambda}~.
\end{equation}
In terms of the boundary length $\ell$, we have
\begin{equation}\label{WdWl}
   \left( \beta^2 \left(\ell \frac{\dd}{\dd \ell} \right)^2  +\frac{\lambda^2 }{4\pi^2 } \, \ell^2\right)\Psi[\ell] = {\varepsilon^2_{\rm m}} \, \Psi [\ell]~.
\end{equation}
Although the wave equation (\ref{HamCyl}) bears a strong resemblance to one that appears when studying quantum gravity with $\Lambda>0$ coupled to a CFT with $c\le 1$ (as reviewed for example in \cite{Ginsparg:1993is}), there are significant differences. In particular, the sign of the kinetic term is opposite and more reminiscent of the situation found in the Wheeler-DeWitt equation for higher dimensional gravity \cite{deWitt}. The `wrong' sign in the kinetic term of the Wheeler-DeWitt equation is often attributed to the existence of a cosmological time in an otherwise `timeless' equation. 
\newline\newline
To render the equation slightly simpler, it is natural to consider the rescaled variables $\tilde{\varphi}_0 = \beta \varphi_0$~, $\varepsilon_{\text{m}} = \beta \tilde{\varepsilon}_{\text{m}}$, and ${\lambda} = \beta \tilde{\lambda}$, and subsequently shift $\tilde{\varphi}_0 \to \tilde{\varphi}_0 - \log \tilde{\lambda}$. The wave equation then reads
\begin{equation}\label{schro}
\biggl( - \frac{\dd^2}{\dd\tilde{\varphi}_0^2} - e^{2 \tilde{\varphi}_0}  \biggr) \Psi[\tilde{\varphi}_0]= -  \tilde{\varepsilon}_{\text{m}}^2 \,  \Psi[\tilde{\varphi}_0]~, 
\end{equation}
The wavefunctions can be found explicitly in terms of Bessel functions \cite{CarneirodaCunha:2003mxy}. The general solution reads
\begin{equation}\label{psi}
 \Psi[\tilde{\varphi}_0] = \alpha_+ \, J_{+\tilde{\varepsilon}_{\text{m}}}\left(  e^{\tilde{\varphi}_0}\right)+ \alpha_- \, J_{-\tilde{\varepsilon}_{\text{m}}}\left(  e^{ \tilde{\varphi}_0} \right)~.
\end{equation}
We now consider separately the cases of positive and negative $\varepsilon_{\text{m}}^2$. 
\newline\newline
\text{\textbf{The case of $\boldsymbol{\varepsilon^2_{\text{m}} > 0}$.}} For the case of $\varepsilon^2_{\text{m}} > 0$ the wave-function with $\alpha_- = 0$ decays to zero as $\varphi_0 \to -\infty$ and oscillates rapidly at large $\varphi_0$. The $\alpha_-$  branch of the wavefunction instead diverges exponentially at large and negative values of $\varphi_0$, or equivalently for arbitrarily small spatial size.  We will set $\alpha_-=0$ for $\varepsilon_{\text{m}}^2>0$ in what follows. The asymptotic behaviour of the $\alpha_+$ branch is given by
\begin{eqnarray}
\lim_{\tilde{\varphi}_0 \to -\infty}  \Psi[\tilde{\varphi}_0]  &\approx& \alpha_+ \left( \frac{1}{2} \right)^{\tilde{\varepsilon}_{\text{m}}} \frac{e^{\tilde{\varepsilon}_{\text{m}} \tilde{\varphi}_0}}{\Gamma(1+\tilde{\varepsilon}_{\text{m}})}~,  \label{largephi0} \\
\lim_{\tilde{\varphi}_0 \to +\infty}  \Psi[\tilde{\varphi}_0]  &\approx&  \alpha_+ \left( {\frac{2}{ \pi }} \right)^{1/2} e^{-\frac{\tilde{\varphi}_0}{2}} \sin \left(    e^{\tilde{\varphi}_0}-\tfrac{1}{2} \pi  \tilde{\varepsilon}_{\text{m}} +\tfrac{1}{4}\pi \right)~. \label{smallphi0}
\end{eqnarray}
The oscillatory behavior at large $\tilde{\varphi}_0$ is indicative of a WKB-type semiclassical state, and is a linear combination of incoming and outgoing waves. Indeed, this is the regime where the spatial circle is parametrically large, and the semiclassical solution is given by the late-time limit of global dS$_2$ (\ref{gdS2}). We can view the state $\Psi$ in the semiclassical large $\varphi_0$ regime as a linear superposition of an expanding and contracting branch of the bouncing dS$_2$ solutions in (\ref{bounde}). This behaviour is characteristic of the Hartle-Hawking \cite{PhysRevD.28.2960} state in gravity with $\Lambda>0$. One could also take a linear combination of the exponentially growing and decaying branches to construct a purely outgoing state, which would correspond to the wavefunction of Vilenkin \cite{Vilenkin:1986cy}.
\newline\newline
As $\beta \to 0$, the wavefunction peaks around the value $e^{\tilde{\varphi}_0} = \tilde{\varepsilon}_{\text{m}}/\tilde{\lambda}$. 
The width of the peak scales as $\beta^{2/3}$, as follows from an Airy function analysis. 
Viewing $\tilde{\varphi}_0$ as a cosmological time parameter, this peak could be an interesting way to set the initial time. 
\begin{figure}
\begin{center}
\includegraphics[scale=0.37]{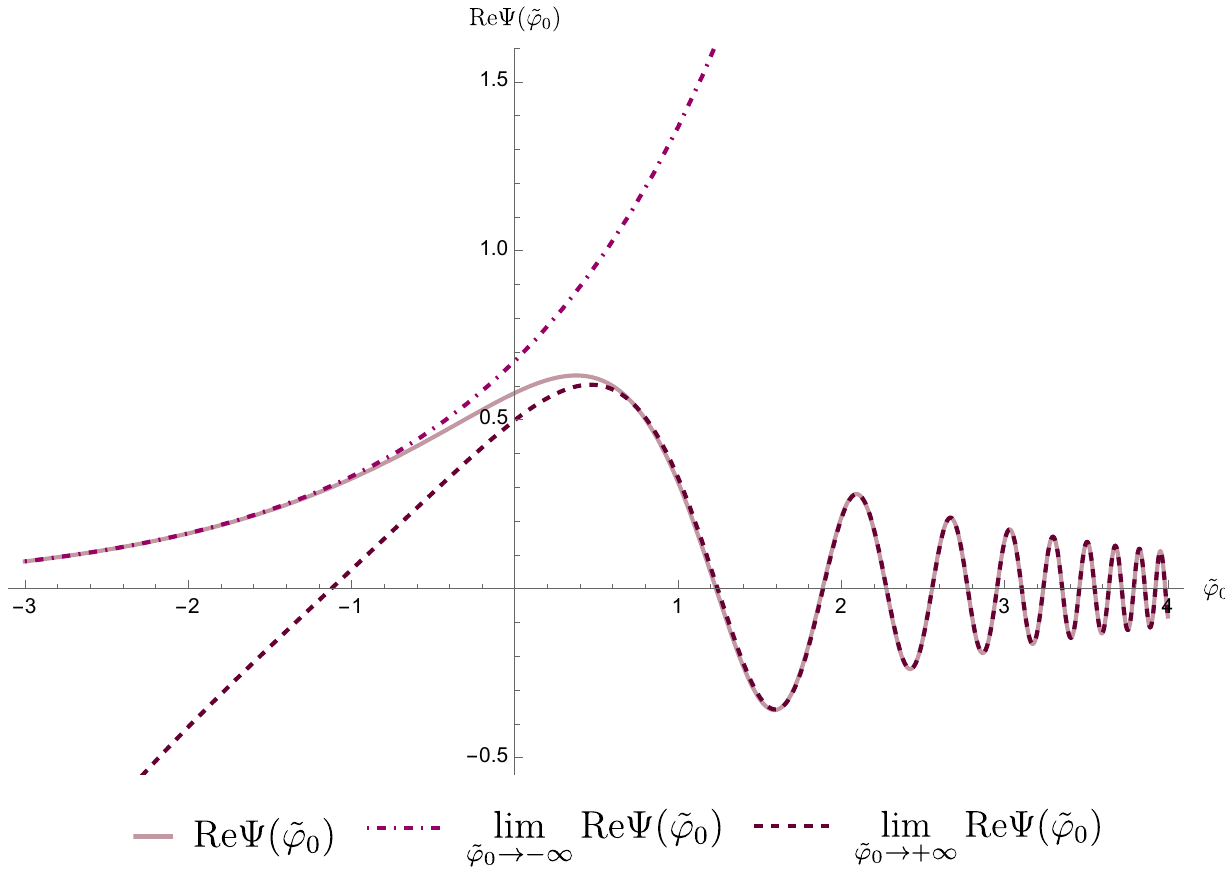}  \includegraphics[scale=0.37]{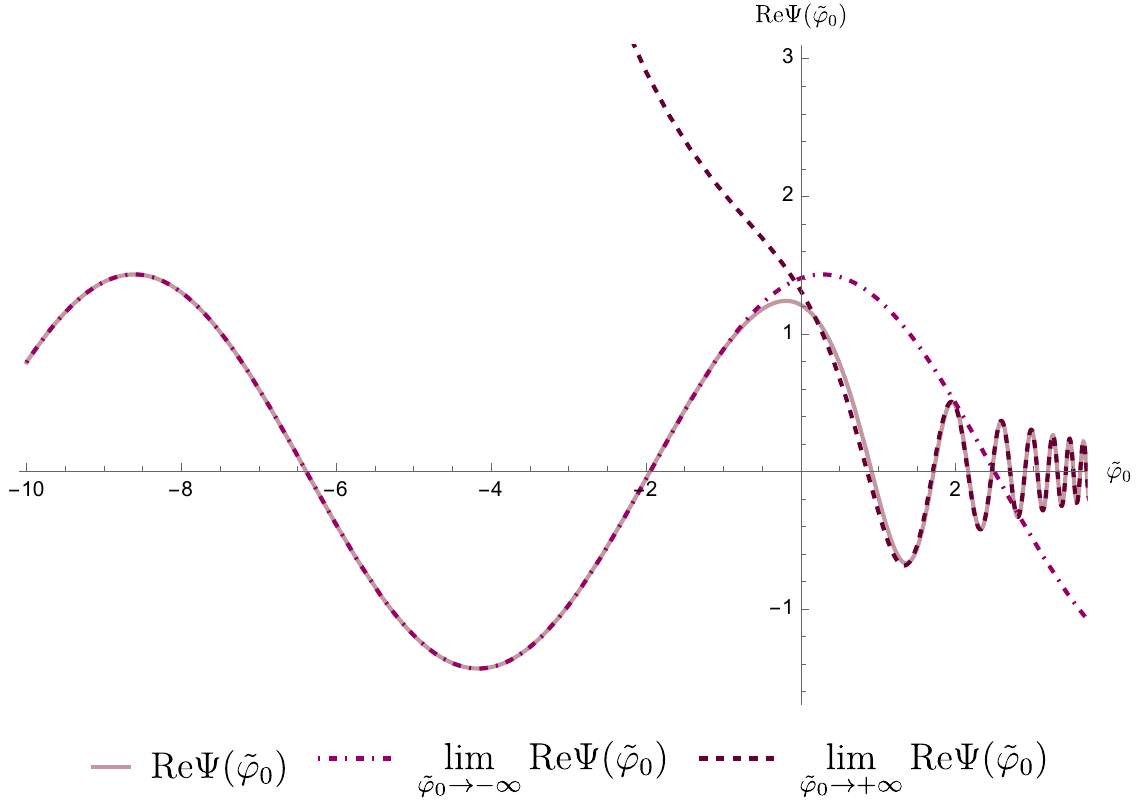}
\end{center}
\caption{\footnotesize{Plot of $\text{Re}  \Psi(\tilde{\varphi}_0)$ vs. $\tilde{\varphi}_0$ in (\ref{psi}) with $\alpha_-=0$ and $\alpha_+=1$ overlaid with the asymptotic approximations (\ref{largephi0}) and (\ref{smallphi0}) for (left) $\tilde{\varepsilon}_{\rm m}^2 = 0.5$ and (right) $\tilde{\varepsilon}_{\rm m}^2 = -0.5$.}}
\end{figure}
$\,$\newline\newline
\text{\textbf{The case of $\boldsymbol{\varepsilon^2_{\text{m}} < 0}$.}} For the case of ${\varepsilon}^2_{\text{m}} < 0$, which occurs when the matter fields are highly excited, the wave function is oscillatory for all $\varphi_0$, but the period of oscillations rapidly grows at large $\varphi_0$.
In particular, for negative values of $\varphi_0$, the wavefunction does not decay to zero. As such, we no longer have a reason to impose that $\alpha_- = 0$ in (\ref{psi}), and we have instead two oscillatory branches for the wavefunction. That the wavefunction remains oscillatory is a quantum mechanical reflection of the fact that one can have classical solutions for arbitrarily small spatial circle in this case, as was seen classically in the big bang/crunch solutions (\ref{bigbang}). In light of this, it is natural to express the wavefunctions such that they are purely outgoing or incoming in the small spatial circle limit. These read
\begin{eqnarray}
\Psi_{\text{in}}[\tilde{\varphi}_0] &=& \alpha_{\text{in}} J_{-i |\tilde{\varepsilon}_{\text{m}}|}\left(  e^{\tilde{\varphi}_0}\right)~, \\
\Psi_{\text{out}}[\tilde{\varphi}_0]  &=& \alpha_{\text{out}} J_{i|\tilde{\varepsilon}_{\text{m}}|}\left(  e^{\tilde{\varphi}_0}\right)~.
\end{eqnarray}
Expanding the above in the limit of large spatial circle size reveals that the purely incoming mode becomes a linear combination of both incoming and outgoing modes, namely
\begin{equation}
\Psi_{\text{in}}[\tilde{\varphi}_0] \approx \alpha_{\text{in}} e^{\tfrac{i\pi}{4}} e^{-\tfrac{\tilde{\varphi}_0}{2}+\tfrac{\pi}{2} |\tilde{\varepsilon}_{\text{m}}|} \left( e^{- i e^{\tilde{\varphi}_0}} - i  e^{-{\pi}| \tilde{\varepsilon}_{\text{m}}|}  e^{+ i e^{\tilde{\varphi}_0} } \right)~.
\end{equation}
A similar expression holds for the modes that are purely outgoing at small spatial circle. This implies that although there are classical solutions that are purely expanding from the big bang to the far future, quantum mechanically there is always a small amplitude, suppressed by
\begin{equation}
e^{-{\pi} |\tilde{\varepsilon}_{\text{m}}|} = e^{-\frac{2 \pi}{\beta}\sqrt{\Delta_{\text{m}}-\tfrac{q^2}{4}-1}}~,
\end{equation}
to have a contracting branch also. We can view this as a quantum mechanical bouncing Universe. A similar feature has been observed in \cite{Hartle_2015,Bramberger_2017} in a minisuperspace approximation of a four dimension model.
\newline\newline
Finally, the case $\varepsilon_{\text{m}} = 0$ is a slightly degenerate case where for large and negative $\varphi_0$ the wavefunctions either tend to a constant or grow linearly in ${\varphi}_0$.

\subsection{Inner-product}

It is interesting to ask whether there is a natural inner product for our wavefunctions. As was previously noted, the wave-equation (\ref{HamCyl1}) has a Klein-Gordon structure. This is reminiscent of the situation in the standard Wheeler-DeWitt formalism \cite{deWitt}. A norm that one could imagine is the Klein-Gordon inner product. In the gauge where $\varphi$ is constant, the Klein-Gordon norm reads
\begin{equation}
(\Psi_1,\Psi_2)_{\text{KG}} =  \frac{i}{2} \left( \Psi_1^* \partial_{\varphi_0}  \Psi_2 - \Psi_2 \partial_{\varphi_0}  \Psi^*_1  \right)~,
\end{equation}
and it is evaluated at some fixed $\varphi_0$. In addition, the norm with respect to the matter degrees of freedom is the standard positive definite quantum field theoretic norm. For a purely expanding $\Psi$, the Klein-Gordon norm is positive, but in general it is not a positive definite norm. In either case, the Klein-Gordon norm has the virtue of being conserved. Interestingly, under this inner-product the states (\ref{psi}), which include the Hartle-Hawking state, have vanishing norm for $\varepsilon_{\text{m}}^2>0$. The states with $\varepsilon_{\text{m}}^2 < 0$ instead can have norm of either sign depending on whether they are dominated by an outgoing or ingoing branch.\footnote{A Klein-Gordon inner product for the Wheeler-DeWitt equation of  three-dimensional general relativity with $\Lambda>0$ has been recently considered in \cite{Araujo-Regado:2022jpj,Godet:2024ich,Verlinde:2024zrh}. Applying it to late-time wavefunctionals, which take the form of a 2d CFT partition function living on $\mathcal{I}^+$, and placing the boundary metric in the Weyl gauge, leads to a putative boundary CFT of central charge $c= 13\pm \tfrac{3i}{2G}$.}
\newline\newline
An alternative norm, at least for a Hartle-Hawking type state, might be given by the sphere path integral. However, this norm would involve integrating over all arguments of the wavefunction, including the scale factor $\varphi$. This is similar to integrating the absolute value squared of a wavefunction over both space and time. Relatedly, along a real contour for $\varphi$ the Euclidean sphere path integral is badly divergent due to the unbounded nature of the timelike Liouville action. To deal with this divergence, one can complexify the contour of integration of $\varphi$, as proposed in \cite{Gibbons:1978ac}. Upon doing so, the resulting sphere path integral becomes complex valued \cite{Polchinski:1988ua,Anninos:2021ene}. As such, the sphere path integral is likely not a good norm unless the theory has non-standard unitarity/Hermiticity properties. It is interesting that the conformal mode problem leads to a complex partition function in Euclidean signature and a non-positive definite inner product in Lorentzian signature. For the case of a spacelike Liouville, where $\varphi$ no longer has a wrong sign kinetic term, it is more natural to integrate over $\varphi$ as well. In turn the sphere path integral of spacelike Liouville, which is real valued, may be a better suited as a choice of norm. 
\newline\newline
Other norms have been proposed in recent literature that involve a Euclidean path integral with more elaborate topologies \cite{Marolf:2020xie,Maxfield:2023mdj,Cotler:2024xzz,Usatyuk:2024mzs}. These norms involve non-trivial topology and are particularly well adapted to topological theories. Such inner products can lead to the surprising feature, at least in sufficiently controlled low dimensional cases, that a closed Universe has a one-dimensional Hilbert space.  In the presence of propagating matter degrees of freedom such norms require a careful treatment due to divergent contributions near small cycle regimes  \cite{Anninos:2022ujl}. Moreover, in more realistic examples, one would also require a method that addresses the conformal mode problem whilst preserving the positivity of the candidate inner product. A recent attempt to overcome some of these issues in the context of AdS$_2$ with matter can be found in \cite{Iliesiu:2024cnh}.


\section{Hartle-Hawking Path Integral}\label{HHPI}

In this section we evaluate, in a saddle point approximation, the Hartle-Hawking path integral \cite{PhysRevD.28.2960} defined by the Euclidean path integral
\begin{equation}\label{HHwf}
\Psi_{\text{HH}} [h,X_{\text{b}}] = \int \mathcal{D}g_{\mu\nu}  \mathcal{D}X \, e^{-S_{\text{tot}}[g_{\mu\nu},X]}~.
\end{equation}
For the case of interest, the action $S_{\text{tot}}$ is the Euclidean action of gravity coupled to matter given in (\ref{Sgrav}), and the matter fields are represented by $X$. The fields reside on a manifold $\mathcal{M}$ with a single boundary $\partial\mathcal{M}$, and we take $\mathcal{M}$ to have disk topology $\mathsf{D}$.\footnote{The latter restriction can be generalised to manifolds of higher genus, and it would be interesting to consider that case also.} One must path integrate over smooth field configurations on $\mathsf{D}$, and must further satisfy a Dirichlet boundary condition fixing the induced metric $h$ and boundary profiles $X_{\text{b}}$ at $\partial \mathsf{D}=S^1$. The wavefunction is constructed so as to solve the Wheeler-DeWitt equation, and more general wavefunctions can be obtained by inserting operators in the gravitational path integral. By comparing our approximate path integral expressions to the exact wavefunctions presented in section \ref{cosmoWF}, we clarify the role of the various saddles. 

\subsection{Conformal gauge}

Once we impose the conformal gauge,
\begin{equation}
ds^2 = e^{2\beta \varphi} \tilde{g}_{\mu\nu} \dd x^\mu \dd x^\nu~,
\end{equation}
and place the matter fields in their Hartle-Hawking state, the non-trivial part of the wavefunction takes the more precise form \cite{David:1988hj,Distler:1988jt}
\begin{equation}\label{hh}
\Psi_{\text{HH}}[\varphi_{\text{b}}] = \frac{1}{\text{vol} \, \text{diff}_{\text{res}}} \int \mathcal{D}\varphi \, e^{-S_{\text{tL}}[\varphi]}~.
\end{equation}
In particular, the measure factor $\mathcal{D}\varphi$ is the standard flat measure on the space of fields $\varphi$, and the overall pre-factor accounts for any remaining residual diffeomorphism redundancies upon fixing the conformal gauge. Finally, we can employ the invariance under tangential diffeomorphisms to fix the Dirichlet condition such that $\varphi_{\text{b}} = \varphi(\rho,\theta)|_{\partial\mathcal{M}}$ takes some constant value $\varphi_0$. We note that the disk path integral of timelike Liouville has been studied from a conformal field theory perspective in \cite{Martinec:2003ka,Bautista:2021ogd}.
\newline\newline
Although the path integral (\ref{hh}) is rather elaborate, we can study it using semiclassical methods \cite{Anninos:2021ene}. As boundary conditions, we impose that the proper size of the boundary circle is fixed to the value $\ell$, and moreover that the boundary conditions preserve the conformal transformation property (\ref{phitransf}). To impose such boundary conditions we must supplement the Euclidean action with a boundary term as follows 
\begin{equation}\label{tSLii}
S_{\text{tL}}[\varphi] =  \frac{1}{4\pi} \int_{\mathsf{D}} \dd^2x \sqrt{\tilde{g}} \left( - \tilde{g}^{\mu\nu} \partial_\mu \varphi \partial_\nu\varphi - q \tilde{R} \varphi + 4\pi \Lambda e^{2 \beta \varphi} \right) -  \frac{q}{2\pi} \int_{S^1} \dd \theta \sqrt{\tilde{h}} \tilde{K} \varphi~.
\end{equation}
The boundary term ensures that the theory is invariant under the redundancies (\ref{phitransf}) associated to the conformal gauge, up to a c-number anomaly  \cite{Fateev:2000ik}. Finally, we must select a particular fiducial metric $\tilde{g}_{\mu\nu}$. 
Here, we take this to be the standard metric on the Euclidean plane 
\begin{equation}\label{cyl}
d\tilde{s}^2 = \dd\rho^2 + \rho^2 \dd \theta^2~,
\end{equation}
with $\theta \sim \theta+2\pi$. We can take the boundary of our manifold to reside at $\rho = 1$.
\newline\newline
Although the path integral we consider is Euclidean, one must generally allow for complex valued saddles. Moreover, the unboundedness of the timelike Liouville action (\ref{tSL}) requires us to deform the original contour of $\varphi$ to one in the space of complexified $\varphi$. As such, the wavefunction can acquire oscillatory behaviour and other features.

\begin{figure}[ht]
\centering
\begin{tikzpicture}

\draw[thick, name path=A] (0,1.2) to[out=90,in=210] (1.,3);
\draw[thick, name path=B] (-1,1.2) to[out=90,in=-30] (-2.,3);
\draw[thick] (-.5,3) ellipse (1.5 and 0.15);
\def\XR{.5}; 
\def\YR{0.1};
\def\xR{.5}; 
\def\xxR{.75}; 
\draw [dashdotted,domain=0:180, blue,thick] plot ({-.5+\xxR*cos(\x)}, {2.3+\YR*sin(\x)}); 
\draw [domain=0:-180, blue,thick] plot ({-.5+\xxR*cos(\x)}, {2.3+\YR*sin(\x)}); 
\draw [dashdotted,domain=0:180,red,thick] plot ({-.5+\XR*cos(\x)}, {1.2+\YR*sin(\x)}); 
\draw [domain=0:-180,red,thick] plot ({-.5+\XR*cos(\x)}, {1.2+\YR*sin(\x)}); 
\draw [domain=0:-180, thick, densely dashed] plot ({-.5+\xR*cos(\x)}, {1.2+\xR*sin(\x)});
\end{tikzpicture}
\caption{A complexified saddle for the Hartle-Hawking no boundary wavefunction can be obtained by gluing the Euclidean dS$_2$ hemisphere solution to the future half of Lorentzian dS$_2$. The saddle caps off smoothly in the Euclidean regime.}
\end{figure}
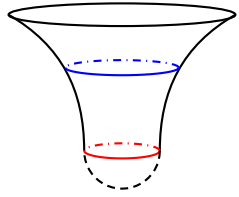

\subsection{Saddle point approximation} 

At small values of $\beta$, we can employ the saddle point approximation. One way to proceed, as was recently done in a related context in \cite{Chaudhuri:2024yau}, is to compute the solution to the saddle point equations and subsequent corrections from nearby fluctuations. The timelike Liouville equation of motion is given by
\begin{equation}\label{seom}
2\tilde{\nabla}^2\varphi - q\tilde{R}  + 8\pi \beta \Lambda e^{2\beta\varphi} = 0~.
\end{equation}
For the fiducial metric (\ref{cyl}) we have that $\tilde{R}=0$, and the above equation simplifies to the ordinary Liouville equation whose general solution is given by an arbitrary holomorphic and anti-holomorphic map of the Fubini-Study metric
\begin{equation}\label{saddle}
\varphi(z,\bar{z}) = \frac{1}{2\beta}  \log \frac{4 f'(z) g'(\bar{z})}{\left(1+ 4\pi \beta^2 \Lambda f(z)g(\bar{z})\right)^2}~,\quad 
\end{equation}
with $z=\rho e^{i \theta}$ and $\bar{z} = \rho e^{-i \theta}$. The functions $f(z)$ and $g(\bar{z})$ are generally allowed to be meromorphic functions. For a real valued Euclidean physical metric, we must further require that $\varphi(z,\bar{z})$ is itself real up to a multiple of $\pi i/\beta$. This can be achieved, for instance, by taking $g(\bar{z}) = \overline{f(z)}$. The induced metric at the boundary $\rho = 1$ is given by
\begin{equation}
h = e^{2 \beta \varphi}|_{\rho=1}~.
\end{equation}
From the above, we note that there is an obstruction in making the physical boundary arbitrarily large whilst retaining a real valued Euclidean bulk metric. When we are interested in large boundary circles, we must release the reality  condition of the Euclidean saddle. 
\newline\newline
The simplest smooth saddle point solution is given by taking $f(z)=\rho_0 z/2 \beta \sqrt{\pi  \Lambda}$ and $g(\bar{z})=\rho_0 \bar{z}/2 \beta\sqrt{\pi  \Lambda}$, such that 
\begin{equation}\label{saddlerho}
e^{2\beta\varphi_*(\rho,\theta)} = \frac{\rho^2_0}{\pi \beta^2 \Lambda} \frac{1}{\left(1+ \rho_0^2 \rho^2 \right)^2}~.
\end{equation}
The above physical metric is the Fubini-Study metric on a portion of $S^2$. One of the poles of the two-sphere resides at $\rho=0$ where the space caps off smoothly, in accordance with the Hartle-Hawking condition. At $\rho =1$ we have an $S^1$ boundary of physical size
\begin{equation}\label{saddleHH}
\ell = \frac{2\pi\rho_0}{\sqrt{\pi \Lambda} \beta } \frac{1}{\left(1+ \rho_0^2 \right)}~, \quad\quad \rho_0 = \frac{ \sqrt{\pi} \pm \sqrt{\pi - \beta^2  \ell^2 \Lambda  } }{   \beta  \ell \sqrt{\Lambda} }~.
\end{equation}
The two roots correspond to a larger ($+$) or smaller ($-$) portion of the two-sphere. The scalar curvature of the physical metric associated to the saddle point solution is $R = 8 \pi \Lambda \beta^2$, such that the physical two-sphere becomes parameterically large at small $\beta$. The range of $\ell$ is bounded by the maximal circle of the $S^2$, and hence cannot become arbitrarily large. Larger values can be attained by dropping the reality condition on the saddle. 
\newline\newline
We can compute the tree-level contribution to the Hartle-Hawking path integral (\ref{hh}) by computing the on-shell action for the saddle (\ref{saddlerho}). To leading order at small $\beta$, with $\Lambda \beta^2$ fixed, we find
\begin{equation}\label{onshell}
\beta^2 S^{\pm}_{\text{tL}}[\varphi_*] = \pm \sqrt{1 -\frac{\beta^2  \ell^2 \Lambda}{\pi}  }-\log \left(1\pm\sqrt{1-\frac{\beta^2 \ell^2 \Lambda  }{\pi }}\right)+\log  \beta^2 \ell \Lambda   +1~.
\end{equation}
%
%
%
%
where the $\pm$ indicates the $\pm$ root in (\ref{saddleHH}).
\newline\newline
It is interesting to compare the above on-shell action to the saddle point approximation of the  wavefunctions (\ref{psi}). We can express the Bessel function in a {Schl\"{a}fli-Sommerfeld} integral form\footnote{See for instance expression 10.9.17 in \href{https://dlmf.nist.gov/10.9}{https://dlmf.nist.gov/10.9}.}
\begin{equation}
J_{\frac{s}{\beta^2}}\left(\frac{z}{\beta^2}\right) = \frac{1}{2\pi i} \int_{\mathcal{C}} \text{d} t \, \exp \, {\frac{\mathcal{I}(t)}{\beta^2}}~,  \quad\quad \mathcal{I}(t) \equiv z \sinh t - s t~,
\end{equation}
where the contour $\mathcal{C}$ goes from $t=+\infty-i\pi$ all the way to $+\infty+i\pi$. The parameter $s \in \{ \pm 1 \}$. In the small $\beta$ limit, whilst keeping $z$ fixed, we can employ a saddle point method. The critical points of $\mathcal{I}(t)$ are given by
\begin{equation} \label{tpm}
t_n^{\pm,s} = \pm s \log\left(1+\sqrt{1-z^2}   \right) \mp s \log z + i  \left(2n+\frac{s-1}{2}\right)\pi~, \quad\quad n \in\mathbb{Z}~.
\end{equation}
Using the above we evaluate 
\begin{equation}\label{sgeneric}
\mathcal{I}\left(t_n^{\pm,s}\right) = \pm \left( \sqrt{1-z^2}-\log \left(1+\sqrt{1-z^2}\right)+\log z\right) - i s \left(2n+\frac{s-1}{2}\right) \pi~.
\end{equation}
For $s=+1$ the appropriate saddle is $t^{+,+}_{n=0}$, since the relevant Bessel function vanishes as $z^{1/\beta^2}$ in the limit $z \to 0^+$ and is real valued. Upon identifying $z = \ell\beta \sqrt{\Lambda/\pi}$, we can compare the $s=+1$ saddle $\mathcal{I}(t^{+,+}_{n=0})$ to the on-shell action $- \beta^2 S^-_{\text{tL}}$ in (\ref{onshell}) and find agreement in the $\ell$-dependence. The $\ell$-dependence of the on-shell action $- \beta^2 S^+_{\text{tL}}$, instead, agrees with $\mathcal{I}(t^{-,+}_{n=0})$ modulo factors of $i \pi$.
\newline\newline
One could proceed along these lines and retrieve subleading corrections in an attempt to get an expression that becomes increasingly close  to the exact expression (\ref{psi}). Such an approach was taken for the spacelike Liouville case in \cite{Chaudhuri:2024yau}. Below, we will consider a different approach to the path integral that first fixes the area of the physical metric. This will reveal the Bessel functions of (\ref{psi}) from a different perspective. 



\subsection{Fixed area method}\label{fixedarea}

\begin{figure}[ht]
\centering
\begin{tikzpicture}
[decoration={markings,mark=at position 0.6 with {\arrow{>}}}] 
\draw[color=gray,thick] (-3.5,0) -- (3.5,0);
\draw[color=gray,thick] (0,-2.3) -- (0,2.3);
\foreach \x in {0,0.05, 0.1,...,3.5}{
    \draw ({\x},0) node[cross=3pt, very thick] {};}
\draw (2.3,0.3) node[above] {$\mathcal{C}$};
\node(n)[inner sep=2pt] at (-3.5,2) {$\upsilon$};
\draw[color=red,very thick,->] (0,-.3) -- (1.5,-.3);
\draw[color=red,very thick] (0,.3) -- (3.5,.3);
\draw[color=red,very thick] (0,-.3) -- (3.5,-.3);
\draw[line cap=round](n.south west)--(n.south east)--(n.north east);
\draw[color=red,very thick] (0,.3) arc (90:270:.3);
\node[scale=.9] at (3.8,0.15) {$\}\, \varepsilon$};
\end{tikzpicture}
\caption{Schl\"afli integration contour. The integration contour avoids the branch cut along the real axis $\mathbb{R}^+$.}
\label{fig:contourPhat}
\end{figure}
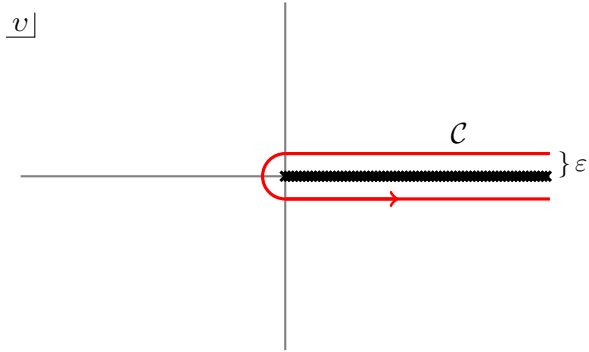

Following \cite{Fateev:2000ik}, let us now consider fixing the area of the physical metric on the disk to some positive real value $\upsilon$. This can be achieved by inserting a delta function 
\begin{equation}
1 = \int_{\mathbb{R}^+} \dd\upsilon \, \delta\left(\upsilon -  \int_{\mathsf{D}} \dd^2x \sqrt{\tilde{g}}\, e^{2\beta\varphi} \right)~, 
\end{equation}
in the path integral. One finds
\begin{equation}
\mathcal{Z}_{\text{disk}}(\upsilon) =  e^{-\Lambda \upsilon} \times  \mathcal{Z}_{\varphi}(\upsilon)~.
\end{equation}
The fixed area timelike Liouville partition function is given by
\begin{multline}\label{diskA}
\mathcal{Z}_{\varphi} (\upsilon)
= \int \mathcal{D}\varphi  e^{-\frac{1}{4\pi} \int_{\mathsf{D}} \dd^2x \sqrt{\tilde{g}} \left( -\tilde{g}^{\mu\nu} \partial_\mu \varphi \partial_\nu\varphi - q \tilde{R} \varphi  \right) +  \frac{q}{2\pi} \int_{S^1} \dd \theta \sqrt{\tilde{h}} \tilde{K} \varphi}  \delta\left(\upsilon -  \int_{\mathsf{D}} \dd^2x \sqrt{\tilde{g}} \,e^{2\beta\varphi} \right) \mathcal{O}_\alpha~,
\end{multline}
where we have allowed for insertion of a timelike Liouville primary 
\begin{equation}
\mathcal{O}_\alpha = \mathcal{V}_{\alpha} \otimes \mathcal{V}_{\text{m}} \otimes \tilde{\mathfrak{c}} \mathfrak{c}~,
\end{equation}
located at the origin of the disk. The reason for including this operator is that timelike Liouville has multiple operators of vanishing scaling dimension \cite{Harlow:2011ny}, and we wish to consider this feature. For the operator $\mathcal{V}_\alpha = e^{2\alpha \varphi}$  to be gauge invariant in the conformal gauge, it must be a conformal primary of vanishing scaling dimension. This is achieved by dressing the operator with the $\tilde{\mathfrak{c}}$ and $\mathfrak{c}$ ghost fields, and a matter primary of scaling dimension $\Delta_{\text{m}}$ satisfying
\begin{equation}\label{eq: total conformal dimension}
\Delta_{\text{m}} + \alpha (q+\alpha) -1 = 0~.
\end{equation}
When $\alpha=0$ and $\alpha=-q$, corresponding to specific non-normaliseable operators (\ref{eq: degenerate fields}), the scaling dimension  (\ref{LDelta}) of $\mathcal{V}_\alpha$ vanishes and we can insert this Liouville operator on its own. In the case of spacelike Liouville theory,  operators related by a reflection such as $\mathcal{V}_\alpha$ and $\mathcal{V}_{-q-\alpha}$ are identified as the same up to an overall normalisation \cite{Teschner:1995yf}. Here, for the non-unitary timelike Liouville case, we will treat them on a separate footing as they are not guaranteed to be equivalent. 
\newline\newline
The saddle point equations of motion are given by 
\begin{eqnarray}
\upsilon &=&  \int_{\mathsf{D}} \dd^2x \sqrt{\tilde{g}}\, e^{2\beta\varphi(\rho,\theta)}~, \label{saddle1} \\
 2 \tilde{\nabla}^2 \varphi(\rho,\theta) &=& -8\pi i \beta \gamma  e^{2\beta \varphi(\rho,\theta)} + \frac{8\pi\alpha}{\rho} \delta(\rho)~,
 \label{saddle2}
\end{eqnarray}
where we have introduced a Lagrange multiplier $\gamma$ for the integral representation of the delta function
\begin{equation} \label{delta}
\delta(x-y) = \int_{\mathbb{R}} \frac{\dd \gamma}{2\pi} \, e^{i \gamma (x- y)}~.
\end{equation}
We now study the above saddle point equations for both vanishing and non-vanishing $\alpha$.
\newline\newline
{\textbf{Case I: No insertion.}} 
In this case there is no insertion in the path integral, and consequently no delta function in the saddle point equation (\ref{saddle2}). A solution to (\ref{saddle1}) and (\ref{saddle2}) is given by the round metric on a portion of the two-sphere with boundary at $\rho=1$:
\begin{eqnarray}
\varphi_* &=& \frac{1}{2\beta} \log \left( \frac{\ell^2\left(1+a^2 \right)^2}{4\pi^2} \frac{1}{(1+ a^2\rho^2)^2} \right)~, \quad\quad a^2 =  \frac{4\pi \upsilon}{\ell^2}-1~, \\
i \gamma_* &=&  \frac{4\pi a^2}{(1+a^2)^2 \beta^2 \ell^2}~.
\end{eqnarray} 
The  area of the physical metric is given by $\upsilon$, whilst the boundary length is $\ell$. 
Recalling the fiducial metric (\ref{cyl}),  we can evaluate the on-shell action for the above solution to leading order at small $\beta$ 
\begin{equation}\label{onshell1}
- \frac{1}{4\pi} \int_{\mathsf{D}} \dd^2 x \sqrt{\tilde{g}} \, \tilde{g}^{\mu\nu}  \partial_\mu \varphi_* \partial_\nu \varphi_* -  \frac{1}{2\pi \beta} \int_{S^1} \dd \theta \, \varphi_* 
= - \frac{1}{\beta^2} \left( \frac{\ell^2}{4 \pi \upsilon }+{ \log \left(\frac{2  \upsilon }{\ell}\right)}-1 \right)~.
\end{equation}
In the above we have used that the boundary $\partial{\mathsf{D}}$ is located at $\rho=1$, that $\tilde{R}=0$, and that $q=1/\beta$ to the order we are working in. We observe that for non-vanishing $\ell$ the action is unbounded from below as $\upsilon$ approaches small positive values. This can be viewed as the fixed area counterpart of the conformal mode problem \cite{Gibbons:1978ac} in Euclidean quantum gravity.
\newline\newline
What remains at this stage is to evaluate the integral over the areas
\begin{equation}\label{schlafli}
\Psi_{\text{HH}}^{(0)}[\ell] \approx  \int_{\mathcal{C}}  \dd\upsilon \left( \frac{2\upsilon}{e\ell}\right)^{\frac{1}{\beta^2}}  \times e^{-\Lambda \upsilon +  \frac{1}{\beta^2} \frac{\ell^2}{4 \pi  \upsilon }}~.
\end{equation}
Again, a saddle point analysis reveals two critical points for $\upsilon$, yielding (\ref{onshell}). More generally, to render the integral (\ref{schlafli}) sensible, the contour $\mathcal{C}$ must be deformed away from the original contour $\upsilon\in\mathbb{R}^+$. A potentially natural choice of contour follows from Schl\"afli's integral representation of the Bessel function. To define the contour, let us take the branch cut of the integrand in the complex $\upsilon$ plane to reside along the positive $\upsilon$ axis. Then the  contour goes around the branch cut by a small imaginary amount $\pm i\varepsilon$ above and below, as denoted in figure \ref{fig:contourPhat}. The contour never touches the origin, and so avoids the essential singularity. To leading order at small $\beta$, with $\beta^2 \Lambda$ fixed, the resulting expression reads
\begin{equation}\label{schlafliA}
\Psi_{\text{HH}}^{(0)}[\ell] \approx \mathcal{N}_\beta \times J_{-\tfrac{1}{\beta^2}}\left( \frac{\ell\sqrt{\Lambda}}{\sqrt{\pi} \beta} \right)~,
\end{equation}
where the normalisation factor $\mathcal{N}_\beta$ is independent of $\ell$. Comparing to (\ref{psi}), we see that the above wavefunction corresponds to the semiclassical limit of the branch which grows at small boundary length. 
\newline\newline
{\textbf{Case II: \boldmath$\alpha=-\eta/\beta$.}} In this case there is a delta function in the saddle point equation (\ref{saddle2}). We are looking for a physical metric of area $\upsilon$, and physical boundary length $\ell$, with positive curvature and a conical singularity at the origin (see for instance \cite{Harlow:2011ny}). We find the following solution
\begin{eqnarray} \label{etasol}
\varphi_* &=& \frac{1}{2\beta} \log \left( \frac{\ell^2\left(1+a^2 \right)^2}{4\pi^2} \frac{1}{ \rho^{4\eta}} \frac{1}{(1+ a^2\rho^{2-4\eta} )^2} \right)~, \quad\quad a^2 =  \frac{4\pi \upsilon(1-2\eta)}{\ell^2}-1~, \\
i \gamma_* &=&  \frac{4\pi a^2}{(1+a^2)^2\beta^2 \ell^2}(1-2\eta)^2~,
\end{eqnarray}
where for the time being we take $\eta < 1/2$, such that $a^2>0$ for sufficiently positive area $\upsilon$ given boundary length $\ell$. To see the conical singularity, we can perform a coordinate transformation $\rho \to \rho^{(1-2\eta)}$ mapping the metric to the standard metric on the two-sphere, but with a non-standard periodicity $\theta \sim \theta+ 2\pi|1-2\eta|$. In our case, it follows from the equations of motion that $\alpha = -\eta/\beta$ to leading order in small $\beta$.
\newline\newline
We must now evaluate the on-shell action at finite area for the above solution. As it is singular, we must exercise caution in evaluating its contribution near the origin where the conical singularity resides. One can follow the procedure of \cite{Zamolodchikov:1995aa} and excise a small disk around the singularity, and renormalise the vertex operator $\mathcal{O}_\alpha$ accordingly to render a finite action. 
In the presence of a heavy operator like the one at hand, the action needs to be slightly modified \cite{Harlow:2011ny} leading to the on-shell action 
\begin{multline}
- \frac{1}{4\pi} \int_{\mathsf{D}\,\backslash \mathsf{D}_{\epsilon}} \dd^2 x \sqrt{\tilde{g}} \, \tilde{g}^{\mu\nu}  \partial_\mu \varphi \partial_\nu \varphi -  \frac{1}{2\pi \beta} \int_{S^1} \dd \theta  \, \varphi  - 2\alpha \varphi(0) + 2 \alpha^2 \log \epsilon 
= \\  - \frac{1}{\beta^2} \left[ \frac{\ell^2}{4 \pi \upsilon }+ (1-2\eta)\left({ \log \frac{2  \upsilon }{\ell}} +\log(1-2\eta)\right) -(1-2\eta) \right]~,
\end{multline}
where $\mathsf{D}_\epsilon$ is a small disk of radius $\epsilon$ carved out at the origin of $\mathsf{D}$.
What remains at this stage is to evaluate the integral over the areas
\begin{equation}\label{eq:eta less12}
\Psi_{\text{HH}}^{(\alpha)}[\ell] \approx  \int_{\mathcal{C}}  \dd\upsilon  \left( \frac{2\upsilon}{\ell}(1-2\eta)\right)^{\frac{1-2\eta}{\beta^2}}  \times e^{-\Lambda \upsilon +  \frac{1}{\beta^2} \frac{\ell^2}{4 \pi  \upsilon } -\frac{1-2\eta}{\beta^2}}~.
\end{equation}
Once again, the original contour of $\upsilon\in \mathbb{R}^+$ fails to yield a convergent answer due to an essential singularity at $\upsilon=0$. To proceed, we again propose a natural choice of contour, as follows from Schl\"afli's integral representation of the Bessel function. Once again, to define the contour let us take the branch cut of the integrand in the complex $\upsilon$ plane to reside along the positive $\upsilon$ axis. Then the  contour goes around the branch cut by a small imaginary amount $\pm i\varepsilon$ above and below. The contour never touches the origin, and so avoids the essential singularity. The resulting expression  reads
\begin{equation}\label{schlafli2}
\Psi_{\text{HH}}^{(\alpha)}[\ell] \approx \mathcal{N}_\beta \times J_{\tfrac{2\eta-1}{\beta^2}}\left( \frac{\ell\sqrt{\Lambda}}{\sqrt{\pi} \beta} \right)~,
\end{equation}
where the normalisation factor $\mathcal{N}_\beta$ is independent of $\ell$. 
\newline\newline
Comparing to (\ref{psi}), we see that upon continuing $\eta$ to the range $\eta > 1/2$, the above wavefunction corresponds to the semiclassical limit of the branch which decays at small boundary length, which is the Hartle-Hawking condition. 
Selecting the value $\eta=+1$, corresponds to inserting a vertex operator $\mathcal{V}_{-q} = e^{- 2 q \varphi}$ of vanishing conformal weight in the Hartle-Hawking path integral.
\newline\newline
Instead of restricting to the regime $\eta <1/2$ with the solution (\ref{etasol}), one can  solve the equation of motion (\ref{saddle2}) for $\eta >1/2$, leading to the solution\footnote{We would like to acknowledge Joel Karlsson for useful discussions on this point.}
\begin{eqnarray} \label{etasol_new}
    \varphi_* &=& \frac{1}{2\beta} \log \left( \frac{\ell^2\left(1+a^2 \right)^2}{4\pi^2} \frac{1}{ \rho^{4\eta}} \frac{1}{(1+ a^2\rho^{2-4\eta} )^2} \right)~, \quad\quad a^2 = \frac{1}{\frac{4\pi\upsilon}{\ell^2}(2\eta-1)-1}~, \\
    i \gamma_* &=&  \frac{4\pi a^2}{(1+a^2)^2\beta^2 \ell^2}(1-2\eta)^2~.
\end{eqnarray}
Slightly adjusting the coefficient of $\log\epsilon$ to account for the divergence near the origin, we obtain the on-shell action for  $\eta >1/2$
\begin{equation}
   S_{\mathrm{tL}}[\varphi_*]=  -\frac{1}{\beta^2}\left[\frac{\ell^2}{4\pi \upsilon} -(2\eta-1)\log \left(\!\frac{2\upsilon}{\ell}(2\eta-1)\!\right) - (2\eta-1) + i\pi(2m+1) \right]\,,~ m \in \mathbb{Z}~.
\end{equation}
We therefore retrieve the same $\upsilon$ and $\ell$ behavior as obtained from the analytic continuation of (\ref{eq:eta less12}) to $\eta >1/2$. 
\newline\newline
As a final remark, it is interesting to contrast the above to what happens in the case of spacelike Liouville theory. There one finds that the disk path integral is given by a modified Bessel function. Concretely, as shown in \cite{Fateev:2000ik,Teschner:2000md}, for an operator insertion $\mathcal{V}_\alpha = e^{2\alpha \varphi}$, one finds that the disk partition function goes as $K_{(Q-2\alpha)/b}(\ell\sqrt{\Lambda}/\sqrt{\pi}b)$. The expression is manifestly invariant under the  reflection map $\alpha \to Q-\alpha$, indicating that one is indeed inserting the same operator when the two operators are related by a reflection. For the case of timelike Liouville theory, one finds instead Bessel functions of the first kind which do not have the same properties as the modified Bessel functions under the respective reflection $\alpha\to -q-\alpha$, leading to the possibility that the two operators with vanishing conformal weight, namely $\mathcal{V}_0$ and $\mathcal{V}_{-q}$, are distinct. Although this cannot occur in a unitary theory, there is no obstruction in a non-unitary theory such as timelike Liouville theory.

\subsection{Brief comment on small fluctuations}

We now briefly discuss the first subleading effect in the saddle point expansion. For simplicity, we discuss the case of $\alpha=0$, for which the saddle is given by (\ref{saddle1}) and (\ref{saddle2}). We are interested in expanding the action in $\varphi = \varphi_* + \delta\varphi$ and $\gamma = \gamma_* +\delta\gamma$ to leading order in the variations. The Liouville field variation $\delta\varphi$ is subject to the Dirichlet condition $\delta\varphi |_{\partial \mathsf{D}} = 0$, as to not disturb the Dirichlet and gauge fixing condition on $\varphi$ at the boundary of the disk. The resulting Gaussian action for $\delta \varphi$  reads
\begin{equation}\label{eq: Gaussian fluctuation}
S^{(2)}[\delta\varphi] = \frac{1}{4\pi} \int_0^{2\tan^{-1}a} d\psi \int_0^{2\pi} d\theta \sin\psi \left( - ( \partial_\psi \delta\varphi )^2 - \frac{1}{\sin^2\psi}( \partial_\theta \delta \varphi)^2 +  2 \delta \varphi^2  \right)~,
\end{equation}
supplemented with the constraint 
\begin{equation} \label{gammaconstraint}
\int_0^{2\tan^{-1}a}  d\psi \int_0^{2\pi}  d\theta  \sin \psi \delta\varphi = 0~.
\end{equation}
For the sake of clarity, we have introduced a coordinate $\tan \tfrac{\psi}{2} = a \rho$ that puts the physical metric in the more standard form
\begin{equation}
ds^2 = \frac{\ell^2 (1+a^2)^2}{16\pi^2 a^2} \left( \text{d} \psi^2 + \sin^2\psi \text{d} \theta^2 \right)~.
\end{equation}
The range of $\psi$ is $(0,2\tan^{-1} a)$. 
\newline\newline
We would now like to path integrate over $\delta \varphi$. This path integral requires complexifying the contour of $\delta\varphi$ to one that is Gaussian suppressed. Upon path integrating over $\delta \varphi$, we obtain a functional determinant for the differential operator
\begin{equation}
\mathcal{O}_a = -\nabla_{S^2}^2 - 2~,
\end{equation}
where $\nabla_{S^2}^2$ is the Laplacian over the round unit two-sphere. The eigenfunctions of eigenvalue $\lambda_a$ which are smooth near $\psi=0$, are given by Legendre functions as follows 
\begin{equation}
\mathcal{Y}_{\lambda, m}(\psi,\theta)  =  e^{i m \theta} \, P_{L_a}^m(\cos \psi ) ~, \quad\quad \text{with} \quad\quad m \in \mathbb{Z}~, 
\end{equation}
where
\begin{equation}
L_a(L_a+1) \equiv  - 2 + \lambda_a~.
\end{equation}
Finally, we must impose that
\begin{equation}
P_{L_a}^m \left(\frac{1-a^2}{1+a^2}\right) = 0~,
\end{equation}
to ensure our Dirichlet condition is satisfied. In all, one is left with a discrete set of $\lambda_a$ that solve the eigenvalue problem. The problem can be solved via numerical means. An efficient approach to do so employs the Gelfand-Yaglom method \cite{Gelfand:1959nq}, which maps the computation to that of a boundary value problem. We leave this analysis to future work. What we would like to note here is that the functional determinant is solely a function of $a^2$ or  equivalently ${\upsilon}/{\ell^2}$. Upon rescaling $\delta\gamma \rightarrow \delta \gamma/\ell^2$, we further note that the higher order terms beyond (\ref{eq: Gaussian fluctuation}) give rise to contributions that solely depend on $a^2$. 

\section{Outlook \& future directions}\label{outlook}

We end with some general remarks and future directions that our analysis invites. At a technical level, a more complete perturbative analysis of the Hartle-Hawking path integral studied in section \ref{HHPI} is of  interest. This may help clarify the precise form of the complexified path integration contours. In turn, this may provide insight into analogous questions for higher dimensional theories. 
\newline\newline
From a more conceptual viewpoint, it is interesting to contrast the global perspective that has been the main focus throughout the text, with a more local static patch perspective.
\newline\newline
\textbf{Cosmological entanglement.} The quantum mechanical condition on the matter theory to be in the big bang/crunch regime discussed in section (\ref{waveeqn}) translates to
\begin{equation}
\Delta_{\text{m}} = \bar{\Delta}_{\text{m}} \ge 1 + \frac{q^2}{4} = \frac{c - 1}{24} ~, 
\end{equation}
as follows from (\ref{mattercond}). In the semiclassical limit, the above scaling dimensions are parametrically large. At sufficiently large $\Delta_{\text{m}}$ we can estimate the number of matter conformal primaries using the Cardy formula
\begin{equation}
\rho(\Delta_{\text{m}},\bar{\Delta}_{\text{m}} ) \approx \exp  2\pi \left(\sqrt{\frac{c}{6}\left( \Delta_{\text{m}} - \frac{c}{24} \right)}+\sqrt{\frac{c}{6}\left( \bar{\Delta}_{\text{m}} - \frac{c}{24} \right)} \right)~.
\end{equation}
If the operators are macroscopically indistinguishable, we can view the above as a microscopic degeneracy associated to the big bang or big crunch singularity. (In light of this, perhaps it is natural to prepare the Universe in a mixed state built from the large collection of such primaries.) As the spacetime expands, and further assuming unitary evolution, the amount of quantum information stored in the particular initial state will be distributed across the future boundary. At sufficiently late times, the spatial slice will have expanded by an exponentially large amount, and is able to store a large amount of quantum information. On the other hand, harboring an arbitrarily complex quantum state on a spatial circle of very small proper size may be in tension with holographic ideas, and is somewhat reminiscent of the idea of a remnant.
\newline\newline
Another proposed measure of quantum information in an asymptotically de Sitter spacetime is the de Sitter entropy $S_{\text{dS}}$ associated to a single static patch of dS$_2$ \cite{Gibbons:1977mu}. This was computed for the theory at hand in \cite{Anninos:2021ene} using Euclidean means. At large $c$, and up to an overall phase, the de Sitter entropy reads
\begin{equation}\label{ZS2}
S_{\text{dS}} = 2\vartheta + \left( \frac{c}{6} - \frac{19}{6} + \ldots \right) \log \frac{c}{\Lambda \ell^2_{\text{uv}}} + \ldots~,
\end{equation}
where $\ell_{\text{uv}}$ is a field theoretic cutoff scale. Consequently, for sufficiently large $\Delta_{\text{m}}$, the amount of quantum information stored in the initial state may lead to a von Neumann entropy on a finite size interval of the late-time slice that exceeds the de Sitter entropy $S_{\text{dS}}$ within a single static patch. Although there is no general expression for the matter entanglement in an excited state, assuming some form of the eigenstate thermalisation hypothesis \cite{Deutsch:1991msp,Srednicki:1994mfb}, we can estimate (see for instance \cite{Asplund:2014coa}) the entanglement in a matter state with $\Delta_{\text{m}}=\bar{\Delta}_{\text{m}} \gg \tfrac{c}{24}$ in an interval of (dimensionless) length $L \in (0,2\pi)$ to be 
\begin{equation}
S_{\text{ent}} \approx \frac{c}{3} \left( \log \sinh\left(\frac{L}{2}\sqrt{\frac{24\Delta_{\text{m}
}}{c}-1}\right) + \log\left( \frac{2 }{\ell_{\text{uv}}\sqrt{\frac{24\Delta_{\text{m}}}{c}-1}} \right) \right)~.
\end{equation}
Using the above expression, we may explore the inequality $S_{\text{ent}} \le S_{\text{dS}}$ as a function of $\Delta_{\text{m}}$. This is a measure of where the matter entanglement begins to compete with the de Sitter entropy of a single static patch. To compare the two, we must carefully select $L$ to reflect the physical size within a dS$_2$ horizon. A violation of such an inequality could indicate a tension somewhat akin to what might occur in the interior of a black hole horizon, thus illustrating the potential limitations of effective field theory. It would be interesting to see how the inclusion of supersymmetry, as in \cite{Anninos:2023exn}, might alter the above expressions.\footnote{It is also interesting to note that in the higher spin model \cite{Anninos:2017eib} of quantum de Sitter space, a vast reduction of the number of degrees of freedom, as compared to the effective field theoretic content, was observed.}
\newline\newline
\textbf{Static patch with timelike boundary.} An alternative physical picture for the disk path integral computed in section \ref{HHPI} is as a thermal partition function for a portion of the static patch in the presence of a timelike boundary. This point of view has been explored recently in \cite{Coleman:2021nor,Anninos:2018svg,Batra:2024kjl,Svesko:2022txo,Banihashemi:2022jys,Anninos:2022hqo,Anninos:2024wpy}, in the hope of sharpening the thermodynamic properties of the de Sitter horizon. From this perspective, the boundary length $\ell$ is interpreted as a physical inverse temperature, and the thermal free energy is given by the logarithm of the disk path integral (\ref{HHwf}) as a function of $\ell$, that is $- \ell F(\ell) = \log \mathcal{Z}_{\text{disk}}(\ell)$. 
\newline\newline
Neither of the two branches of the Wheeler-DeWitt solutions (\ref{psi}), when interpreted as a thermal partition function, admit standard positivity properties. For instance, positivity of the partition function only occurs for restricted temperature ranges. As we lower the temperature the partition function becomes oscillatory, suggesting the appearance of negative norm states in the gravitational sector, echoing ideas in \cite{Anninos:2021ydw}. Computing other thermodynamic quantities using the exact expressions of section \ref{cosmoWF}, reveals that the specific heat $C(\ell) = \ell^2 \partial^2_\ell \log \mathcal{Z}_{\text{disk}}(\ell)$ can be arranged to be positive only at sufficiently high temperatures upon using the $\alpha_-$ branch of the wavefunction. From the on-shell action (\ref{onshell}), we can compute the semiclassical contribution to the thermodynamic entropy $S(\ell) = (1-\ell \partial_\ell)\log\mathcal{Z}_{\text{disk}}(\ell)$. Expanding the semiclassical expression at small $\ell$, we find
\begin{equation}
    S(\ell) = \frac{1}{2\beta^2} \left( \mp \log\beta^2 \ell^2 \Lambda - \log \beta^2 \Lambda + \ldots \right)~,
\end{equation}
It is also worth noting that the positive branch of the entropy $S(\ell)$ receives a contribution logarithmic in $\Lambda$ (but independent of $\ell^2 \Lambda$) whose coefficient is one-half the one for the two-sphere path integral (\ref{ZS2}) computed in \cite{Anninos:2021ene}. This is in line with the fact that the static patch with a boundary retains only one of the two pieces of the dS$_2$ horizon.
\newline\newline
It would be interesting to explore the disk path integral in the supersymmetric counterpart \cite{Anninos:2022ujl,Anninos:2023exn} of the model under consideration. Perhaps the analysis is amenable to supersymmetric index techniques. Finally, given that the fixed area disk partition function explored in \ref{fixedarea} reveals a Euclidean AdS$_2$ saddle, it would seem that more standard AdS/CFT techniques can be at play. 
\newline\newline
\textbf{Topology \& Wheeler-DeWitt.} As a final comment, we emphasise that our analysis of the Hartle-Hawking path integral has been restricted to disk topology. We could also consider the path integral on non-trivial topology, by adding handles to the disk. In this case, the path integral over the matter fields will produce divergences when we integrate over regimes of small cycle size, and the higher-genus contributions may start to compete with the lower genus contributions \cite{Anninos:2022ujl}. We may view this as an indication that although the model is ultraviolet finite, it is not ultraviolet complete. 
\newline\newline
A more controlled setting where we can keep track of these contributions is two-dimensional quantum gravity with $\Lambda>0$ coupled to a $c\le 1$ conformal field theory. There, higher-genus contributions can qualitatively alter the structure of the Wheeler-DeWitt wavefunctions/equation at large $\ell$, as explored in \cite{Banks:1989df,Moore:1991sf,Moore:1991ir,Betzios:2020nry}. Although, naively, the effects are exponentially suppressed in the topological coupling $\vartheta$, they can become significant at sufficiently large $\ell$ \cite{Moore:1991sf,Betzios:2020nry}.\footnote{A gauge theoretic analogue of this effect has been observed in the late-time behaviour of the Schwinger model placed on a dS$_2$ background \cite{Anninos:2024fty}.} From the  cosmological perspective taken in this article, we might imagine such effects (if present) as altering the deep infrared quantum structure of an exponentially expanding spacetime.

\section*{Acknowledgements}

We would like to acknowledge useful and interesting discussions with Tarek Anous, Panagiotis Betzios, Eleanor Harris, Thomas Hertog, Joel Karlsson, Edgar Shaghoulian, Eva Silverstein, Joel Karlsson, and especially Alex Belin. D.A. is funded by the Royal Society under the grant ``Concrete Calculables in Quantum de Sitter" and the STFC Consolidated grant ST/X000753/1. C.B. is funded by STFC under grant number STFCMATH2302.

\bibliographystyle{JHEP}
\bibliography{bib}
\end{document}